\documentclass[11pt]{article}
\usepackage{amssymb}
\author{K. Anguige\\K.P.Tod}
\title{\textbf{Isotropic Cosmological Singularities I.\\Polytropic perfect fluid spacetimes.}}
\begin{document}
\maketitle
\begin{abstract}
We consider the conformal Einstein equations for~$1\leq\gamma\leq
2$~polytropic perfect fluid cosmologies which admit an isotropic
singularity. For~$1<\gamma\leq 2$~it is shown that the Cauchy problem
for these equations is well-posed, that is to say that solutions
exist, are unique and depend smoothly on the data, with data
consisting of simply the
3-metric of the singularity. The analogous result 
for~$\gamma=1$~(dust) is obtained when Bianchi type symmetry is assumed.
\end{abstract}
\section{Introduction}

Isotropic singularities are a class of cosmological singularities
which are both physically interesting and mathematically tractable
(Goode and Wainwright 1985, Tod 1987, Tod 1992). Put simply, an isotropic singularity is one which can be removed by rescaling the spacetime metric with a single function which becomes singular on a spacelike hypersurface.

The motivation for studying these singularities comes from Penrose's
Weyl Tensor Hypothesis (WTH) (Penrose 1979, Penrose 1990). Penrose hypothesised
that initial singularities, notably the Big Bang, should be different
in character from final singularities, as formed in gravitational
collapse or the Big Crunch. Specifically (Penrose 1979), he suggested:
\begin{quote} I propose that there should be a complete lack of chaos in the initial geometry\ldots This restriction on the early geometry should be something like: the Weyl curvature~$C_{abcd}$~vanishes at any initial singularity.
\end{quote} 
Penrose expected this condition to imply that the subsequent evolution
would necessarily be close to a Friedman-Robertson-Walker (FRW)
model. He was led to 
the above hypothesis by the need for some kind of low entropy
constraint 
on the initial state of the universe, at the same time as the matter
content was in thermal equilibrium. He argued firstly that the need for
a low entropy constraint follows from the existence of a second law of
thermodynamics and secondly that low entropy in the gravitational
field must be 
tied to constraints on the Weyl curvature. Penrose went on to suggest
that this constraint on the initial Weyl tensor should be a
consequence of an as yet undiscovered quantum theory of gravity.

In order to make mathematical progress with the WTH one must first say
exactly what is meant by a `finite Weyl tensor' singularity, a
singularity at which the Weyl tensor is finite while, of necessity,
the Ricci tensor is not. There may
be more than one plausible way to do this. One strategy, and the one
we shall adopt, is first to make the
following definition: A spacetime~$(\tilde{M},\tilde{g}_{ab})$~is 
said to admit an~\emph{isotropic singularity}~if there exists a manifold with boundary~$M\supset\tilde{M}$, a regular Lorentz metric~$g_{ab}$~on~$M$, and a function~$\Omega$~defined on~$M$,~such that
\begin{equation}\tilde{g}_{ab}=\Omega^{2}g_{ab}~~~~\textrm{for}~~~\Omega>0\end{equation}
\begin{equation}\Omega\rightarrow 0~~~~\textrm{on}~\Sigma\end{equation}
where~$\Sigma$~is a smooth spacelike hypersurface in~$M$, called the \emph{singularity surface}.
Since the Weyl tensor with its indices arranged as~$C^{a}_{~bcd}$~is 
conformally invariant, isotropic singularities form a well-defined
class of cosmological singularities with finite Weyl tensor.

Consideration of the conformal Einstein equations near an isotropic 
singularity, for various matter models, leads naturally to a class of 
singular Cauchy problems for the unphysical metric~$g_{ab}$, with data
given on~$\Sigma$, for which one may seek to prove suitable existence
and uniqueness theorems. This program was begun in (Tod 1990, 1991), 
where the problem was treated in the case of a perfect 
fluid with polytropic equation of state as matter source. There the 
conformal field equations were
written as an evolution system for~$g_{ab}$~with data just the
3-metric of the singularity surface. It was noted in particular that a
uniqueness result for these equations would prove an earlier
`Vanishing Weyl Tensor' conjecture,(Tod 1987), which said that, 
given an equation of state, the Weyl
tensor must vanish everywhere in~$M$~if it vanishes at an initial
isotropic singularity.
 
Newman (1993a,b) considered cosmologies with an
isentropic perfect fluid as source and for which the conformal 
factor~$\Omega$~remained a smooth coordinate at~$\Sigma$. This 
smoothness condition implied that the equation of state approached 
that for radiation ($\gamma=4/3$) at the singularity. By 
taking~$\gamma=4/3$~throughout~$\tilde{M}$~Newman was able, with a careful 
choice of field variables, to write the equations in symmetric 
hyperbolic form but with a singular~`$\frac{1}{t}$'~forcing term. 
Again the free data at the singularity consisted of just the 3-metric 
there. As there was no existence and uniqueness theorem available for 
the singular evolution equations obtained, a special study was 
undertaken by Newman, the result appearing in (Claudel and Newman
1998). We shall refer to the
existence and uniqueness theorem obtained there as the Newman-Claudel 
Theorem. The conclusion is that there is exactly
one~$\gamma=4/3$~perfect fluid cosmology with an 
isotropic singularity for each 3-metric given on~$\Sigma$. 

Our aim in this paper is to show that by imposing a 
suitable~$\gamma$-dependent differentiability condition on the 
conformal factor~$\Omega$, a similar analysis to that of Newman can be
performed for polytropic perfect fluids with index in the 
range~$1<\gamma\leq 2$. It will then follow from the Newman-Claudel
Theorem that we have existence and uniqueness of
a perfect fluid cosmology for each~$\gamma$ in this range, for each
3-metric on~$\Sigma$.\\

The plan of the paper is as follows:

In section 2 we set down our conventions and give a few notes on 
conformal rescaling.

An account of the theory of singular symmetric hyperbolic systems
developed by Claudel and Newman (1998) is given in section 3.

In section 4 we consider cosmologies  with~$1<\gamma<2$. Using
comoving coordinates as in (Newman 1993a,b) we are able to define a
set of variables in terms of which the conformal field equations can
be written in a suitable singular symmetric 
hyperbolic form. The results of (Claudel and Newman 1998) 
can be applied and the
Newman-Claudel theorem then gives existence and uniqueness of solution.

The case~$\gamma=2$~has to be treated separately and this we do in
section 5. We use harmonic rather than comoving coordinates and this
enables us once again to write the field equations in the form
required in (Claudel and Newman 1998) and to conclude existence 
and uniqueness of
solutions from the Newman-Claudel theorem.

Finally in section 6 we consider the case~$\gamma=1$. Here it has not
been possible to write the conformal equations in full generality in symmetric
hyperbolic form and for this reason we are forced to restrict
attention to spatially-homogeneous solutions with Bianchi type
symmetry. Subject to this
restriction, we are able to prove existence and uniqueness of
solutions, given an initial, spatially-homogeneous 3-metric, by an
extension of a theorem from (Rendall and Schmidt 1991).\\

\section{Conventions and conformal rescaling}
Throughout the paper we take the spacetime metric to have
signature~\\$(+---)$. 
\\For a metric~$g_{ab}$~our definition of the Riemann tensor is 
\begin{equation}(\nabla_{a}\nabla_{b}-\nabla_{b}\nabla_{a})V^{c}=R^c_{~dab}V^{d}\end{equation}
while for the Ricci tensor we take
\begin{equation}R_{ab}=R^c_{~acb}\end{equation}
\\Now whenever we're considering a rescaling as in (1)-(2), tilded quantities will always refer to the singular, physical spacetime $(\tilde{M},\tilde{g}_{ab})$, while un-tilded quantities will refer to the regular, unphysical spacetime $(M,g_{ab})$. 
For metrics $g_{ab}$ and $\tilde{g}_{ab}$ related by (1), the Ricci tensors are related by:
\[\tilde{R}_{ab}=R_{ab} - 2\nabla_{a}\nabla_{b}\log\Omega+2\nabla_{a}\log\Omega\nabla_{b}\log\Omega\] \begin{equation}- g_{ab}(\square\log\Omega+2\nabla_{c}\log\Omega\nabla^{c}\log\Omega)\end{equation}
and the Weyl tensors by:\begin{equation}\tilde{C}^{a}_{~bcd}=C^{a}_{~bcd}\end{equation}
If $\tilde{t}^{a}$ is a unit vector in $\tilde{M}$ then
$t^{a}=\Omega\tilde{t}^{a}$ is one in $M$, and if $\tilde{w}_{a}$ is a
unit covector in $\tilde{M}$, then $w_{a}=\Omega^{-1}\tilde{w}_{a}$ is
one in $M$. With respect to a choice of unit timelike vector $t^{a}$,
the electric and magnetic parts of the Weyl tensor of~$g_{ab}$~are
defined, respectively,
by:\begin{equation}E_{ab}=C^{c}_{~adb}t_{c}t^{d} \qquad
B_{ab}=\frac{1}{2}\epsilon_{ae}^{~~cd}C^{f}_{~bcd}t^{e}t_{f}\end{equation}
The projection~$h_{ab}$~orthogonal to ~$t^{a}$~is defined by
\begin{equation}h_{ab}=g_{ab}-t_{a}t_{b}\end{equation}
If~$t^{a}$~is hypersurface orthogonal one then calculates that
\begin{displaymath}E_{ab}=-{}^{3}R_{ab}+KK_{ab}-K_{ae}K^{e}_{~b}+\frac{1}{2}h_{a}^{~r}h_{b}^{~s}R_{sr}\end{displaymath}
\begin{equation}+\left(\frac{1}{2}h^{cd}R_{cd}-\frac{1}{3}R\right)h_{ab}\end{equation}
where~${}^{3}R_{ab}$~is the Ricci tensor
of~$h_{ab}$,~$K_{ab}$~is the
second fundamental form of the surfaces orthogonal to~$t^{a}$~and ~$K=h^{ab}K_{ab}$.
\\Also
\begin{equation}B_{ab}=D_{d}K_{c(b}e_{a)}^{~~cd}\end{equation}
where~$D_{a}$~is the intrinsic covariant derivative associated
with~$h_{ab}$~and~$e_{abc}$~is the volume element associated with~$h_{ab}$.
\\By (6)-(7) we have 
\begin{equation}\tilde{E}_{ab}=E_{ab} \qquad \tilde{B}_{ab}=B_{ab}\end{equation}
\noindent It follows from (1) that, if $\Omega$ vanishes at a surface
$\Sigma$ in $M$ then $\Sigma$ is a singularity in $\tilde{M}$ but,
from (11), 
$\tilde{E}_{ab}$ and $\tilde{B}_{ab}$ are finite at the
singularity. It follows from (10) that if $\Sigma$ has vanishing second
fundamental form $K_{ab}$ then $B_{ab}$ is automatically zero at
$\Sigma$ and from (9), subject to the term in the 4-dimensional Ricci
tensor being a multiple of $h_{ab}$, that $E_{ab}$ also vanishes at $\Sigma$ if and only
if $\Sigma$ is a 3-dimensional space of constant curvature. This will
be significant in section 5.5.

\section{Singular evolution equations}
The fundamental technique used by Newman (1993b) in tackling the conformal 
Cauchy problem for the Einstein-perfect fluid equations was to choose 
certain field variables so that the resulting equations appeared in 
symmetric hyperbolic form (as in equation (15) below). 
Although the system was
hyperbolic, the unavoidable~`$\frac{1}{t}$'~singularity in these
equations precluded the use of standard existence and uniqueness 
theorems as found in e.g. (Kato 1975, Racke 1992). A special study was
undertaken by Newman to deal with the singularity, and he was able 
to prove well-posedness of the Cauchy problem for a class of singular 
evolution equations into which the~$\gamma=4/3$~equations fell
(Claudel and Newman 1998).  

We will show in section 4 that the conformal field equations 
for other~$\gamma$~in the range~$1<\gamma\leq 2$~can, after a little 
algebra, be forced into the class studied in (Claudel and Newman
1998), giving us 
the result we are looking for.

For ease of reference we now present the main theorem of (Claudel and
Newman 1998), both in its abstract form and its application 
to symmetric hyperbolic systems.\\

\textbf{Theorem 1~(Newman-Claudel~1998).}
Let
\begin{equation}X^{2q+1}\hookrightarrow\ldots\hookrightarrow
X^{1}\hookrightarrow X^{0}\end{equation}
be a sequence of continuous dense inclusions of reflexive Banach
spaces, and let ~$X\equiv X^{0}$. 
\\ Consider the quasi-linear, first
order evolution problem
\begin{displaymath}A^{0}_{t}(u_{t})\frac{d}{dt}u_{t}=A_{t}(u_{t})u_{t}+f_{t}(u_{t})+\frac{1}{t}F_{t}(u_{t})u_{t},~~~0\leq
t\leq T,\end{displaymath}
\begin{equation}u(0)=u^{0}\end{equation}
in an open neighbourhood~$U$~of~$u^{0}\in X$~for
families~$\{f_{t}(x)\in X\}_{(t,x)\in [0,T]\times U}$,\linebreak~$\{F_{t}(x)\in
B(X)\}_{(t,x)\in [0,T]\times U}$.~($B(X)$~being the space of bounded
operators in~$X$),~and a family~$\{A_{t}(x)\}_{(t,x)\in [0,T]\times
U}$~of operators in~$X$ such that\\
~~(i)~~$X^{p+1}\subset D((A_{t}(x))^{p})$~for all~$(t,x)\in [0,T]\times
U$;\\
~~(ii)~~$u^{0}\in\bigcap_{i=0}^{2q+1}X^{p}$;\\
~~(iii)~~$F_{t}(x)u^{0}=0$~for all~$(t,x)\in [0,T]\times U$.
\\\\
Let~$U^{0}\subset U$~be an open neighbourhood
of~$u^{0}$~in~$X^{0}$~and, for each~$p\geq 0$,~let~$U^{p+1}$~be an
open neighbourhood of~$u^{0}$~contained in~$U^{p}$. Suppose there
exists~$\beta\geq 0$~and, for each~$(t,x)\in [0,T]\times U$, an
equivalent renorming~$X_{(t,x)}$~of~$X$, such that\\
\\~~(iv)~~$A_{t}(y)\in G(X_{(t,y)},1,\beta)$~for all~$(t,y)\in
[0,T]\times U^{1}$.~(i.e~$A_{t}(y)$~generates a
contraction~$C_{0}$~semigroup on~$X_{(t,y)}$);\\
~~(v)~~$[0,T]\times U\ni x\mapsto\|~\|_{X_{(t,x)}}$~is Lipschitz
continuous;\\
~~(vi)~~for each~$t\in [0,T]$~and all~$p\geq 0,~U^{p}\ni x\mapsto
f_{t}(x)$~is in~$Lip(U^{p},X^{p})$,~$U^{p}\ni x\mapsto
F_{t}(x)|X^{p}$~is in~$Lip(U^{p},B(X^{p}))$~and~$U^{p}\ni x\mapsto
A_{t}(x)|X^{p+1}$~is in\linebreak $Lip(U^{p},B(X^{p+1},X^{p}))$, with~$t$-independent Lipschitz constants in each case;\\
~~(vii)~~for all~$p,~r:~0\leq r\leq p$~one has~$f|[0,T]\times
U^{p-r}\in C^{r}([0,T]\times U^{p-r},X^{p-r})$, and that~$[0,T]\times
U^{p-r}\ni (t,x)\mapsto F_{t}(x)|X^{p-r}$~is in~$C^{r}([0,T]\times
U^{p-r},B(X^{p-r}))$, and that~$[0,T]\times U^{p-r}\ni (t,x)\mapsto
A_{t}(x)|X^{p-r+1}$~is in\linebreak~$C^{r}([0,T]\times U^{p-r},B(X^{p-r+1},X^{p-r}))$.\\\\
Suppose there exists a linear subspace~$K$~of~$X$, contained and dense
in ~$X^{p}$~for all~$p\geq 0$, and suppose that for all~$p\geq
0$~there exist~$S^{(p)}\in C^{1}([0,T],B(X^{p+1},X^{p}))$~and
operators~$\{E_{t}^{(p)}(x)\}_{(t,x)\in [0,T]\times
U^{p+1}}$~in~$X$~such that\\\\
~~(viii)~~$S_{t}^{(p)}:X^{p+1}\rightarrow X^{p}$~is an isomorphism for
each~$p$, and an isometry at~$t=0$;\\
~~(ix)~~$\frac{d}{dt}S_{t}^{(p)}x'=[A_{t}(x),S_{t}^{(p)}]x'+E_{t}^{(p)}(x)S_{t}^{(p)}x'$~for
all~$x'\in K$~and all~$(t,x)\in [0,T]\times U^{p+1}$;\\
~~(x)~~$[0,T]\times U^{p+1}\ni (t,x)\mapsto
E_{t}^{(p)}(x)$~is~$B(X^{p})$-norm bounded.\\
\\Suppose also that there exists an integer~$q\geq 2$~such that\\\\
~~(xi)~~the spectrum of~$(A^{0}_{0})^{-1}F_{0}(u^{0})\in B(X)$~contains
no integer in the interval~$[1,q]$;\\
~~(xii)~~$\|((A^{0}_{0})^{-1}F_{0})(u^{0})\|_{B(X^{p})}< q-1$~for
all~$p: 0\leq p\leq q$.\\\\
Lastly suppose that~$A^{0}: [0,T]\times U\rightarrow B(X)$~satisfies\\
~(xiii)~~$A^{0}_{t}(x)$~has range~$X$~for all~$(t,x)\in [0,T]\times
U$;\\
~~(xiv)~~there exists~$\alpha >0$~such that~$\alpha I-A^{0}_{t}(x)$~is
dissipative on~$X_{(t,x)}$~for all~$(t,x)\in [0,T]\times U$.~(i.e
$\|(A^{0}_{t}+(\lambda-\alpha)I)x\|_{X_{(t,x)}}\geq\lambda\|x\|_{X_{(t,x)}}~\forall
\lambda\geq 0)$;\\
~~(xv)~~$A^{0}_{t}(x)$~leaves~$X^{p}$~invariant for all~$(t,x)\in
[0,T]\times U$~and all~$p\geq 0$;\\
~~(xvi)~~$U^{p}\ni x\mapsto A^{0}_{t}(x)|X^{p}$~is
in~$Lip(U^{p},B(X^{p}))$~for all~$t\in [0,T]$~and all~$p\geq 0$;\\
~(xvii)~~$[0,T]\times U^{p}\ni (t,x)\mapsto A^{0}_{t}(x)|X^{p}$~is
in~$C^{p}([0,T]\times U^{p},B(X^{p}))$~for all~$p\geq 0$;\\
~(xviii)~~$[0,T]\times U^{p+1}\ni (t,x)\mapsto
[(A^{0}_{t}(x))^{-1},S_{t}^{(p)}]A_{t}(S_{t}^{(p)})^{-1}$~is~$B(X^{p})$-norm
bounded for all~$p: 0\leq p\leq q-1$.\\\\
Let
\begin{equation}U^{0}=\{(u^{0})'\in U: F_{t}(x)(u^{0})'=0~\forall
(t,x)\in [0,T]\times U\}\end{equation}
Then there exists~$T_{0}\in (0,T]$~and an open
neighbourhood~$\hat{U}^{2q+1}\subset
U^{2q+1}$~of~$u^{0}$~\linebreak in~$X^{2q+1}$~such that, for each~$(u^{0})'\in
U^{0}\cap \hat{U}^{2q+1}$~the evolution problem (13) has a unique
solution~$u'\in \bigcap_{p=0}^{q}C^{p}([0,T_{0}],U^{q-p})$, and the
mapping~$U^{0}\cap \hat{U}^{2q+1}\ni (u^{0})'\mapsto u'\in
C^{0}([0,T_{0}],U)$~is continuous with respect to
the~$X^{2q+1}$\linebreak[4] topology on~$U^{0}\cap\hat{U}^{2q+1}$.
\\\\
By taking~$X^{p}=H^{p+s}(\mathbb{R}^{m},\mathbb{R}^{N})$~,
where~$H^{r}(\mathbb{R}^{m},\mathbb{R}^{N})$~is the ~$L^{2}$-type
Sobolev space (Adams 1975) of ~$\mathbb{R}^{N}$-valued functions
on~$\mathbb{R}^{m}$, and~$s>m/2$, Newman-Claudel are able to apply the
above theorem to symmetric hyperbolic PDE, yielding the following
result:\\\\

\textbf{Theorem 2~(Newman-Claudel~1998).} 
Consider the system of
quasi-linear PDE
\begin{displaymath}A^{0}(u_{t})\partial_{t}u_{t}=A^{j}(u_{t})\partial_{j}u_{t}+f(u_{t})+\frac{1}{t}F(u_{t})u_{t},~~0\leq
t\leq T\end{displaymath}
\begin{equation}u(0)=u^{0}\end{equation}
on~$\mathbb{R}^{m}$~for an unknown function~$u: [0,T]\mapsto
U$,~where~$U$~is an open neighbourhood
of~$u^{0}$~in~$X=H^{s}(\mathbb{R}^{m},\mathbb{R}^{N}),~f$~is
an~$X$-valued function on~$U$, and~$A^{0},\ldots
,A^{m},F$~are~$B(X)$-valued functions on~$U$.\\\\
Assume
\\~~~(1)~~ $A^{0}(x),\ldots ,A^{m}(x),F(x)$~have the form of~$N\times
N$~matrices, and~$f(x)$~has the form of an~$N$-component column vector
with each matrix or column element a rational function
(over~$\mathbb{R}$) of the column elements\linebreak $x_{1},\ldots
,x_{N}$~of~$x$;\\
~~~(2)~~ $A^{0}(x),\ldots ,A^{m}(x)$~are symmetric matrices.
\\\\Assume also that the initial data~$u^{0}$~satisfies\\
~~~(3)~~$F(x)u^{0}=0$~for all~$x\in U$;\\
~~~(4)~~ the range of~$u^{0}$~in~$\mathbb{R}^{N}$~is uniformly bounded
away from the zero sets of the denominators in (1);\\
~~~(5)~~ $A^{0}(u^{0})$~is uniformly positive definite as an~$N\times
N$~real matrix-valued function on~$\mathbb{R}^{m}$;\\
~~~(6)~~$u^{0}\in\cap_{r=0}^{2q+1}H^{r}(\mathbb{R}^{m},\mathbb{R}^{N})$.\\\\
Finally assume there exists an integer~$q\geq 2$~such that\\\\
~~~(7)~~for all~$\mathbf{\xi}\in\mathbb{R}^{m}$~the
matrices~$(A^{0})^{-1}F(u^{0}(\mathbf{\xi}))$~have no integer
eigenvalues in the interval~$[1,q]$;\\
~~~(8)~~$\|(A^{0})^{-1}F(u^{0}(\mathbf{\xi}))\|_{B(\mathbb{R}^{N})}<q-1$~holds
for~$\xi\in\mathbb{R}^{m}$.\\\\
Then for
initial data~$u^{0}\in H^{2q+1+s}(\mathbb{R}^{m},\mathbb{R}^{N})$,
there exists a unique
solution~$u\in\cap_{p=0}^{q}C^{p}([0,T_{0}],H^{q-p+s}(\mathbb{R}^{m},\mathbb{R}^{N}))$~of
(15), with an\linebreak $H^{2q+1+s}(\mathbb{R}^{m},\mathbb{R}^{N})$-small
change in the initial data~$u^{0}$~giving rise to
a\linebreak $C^{0}([0,T_{0}],H^{s}(\mathbb{R}^{m},\mathbb{R}^{N}))$-small change
in~$u$.

\section{The Cauchy problem for polytropic perfect fluid spacetimes}
In this section we consider the conformal Einstein equations for
polytropic perfect fluid spacetimes which admit an isotropic
singularity, extending the work of Tod (1990, 1991) and
Newman (1993a,b). The hard work is to find suitable
variables to put a reduced form of the equations into the form (15),
and to see that a solution of the reduced equations does give a
solution of the Einstein equations. The conclusion will be that 
for~$1<\gamma<2$~the conformal
Cauchy problem for the conformal Einstein equations is well posed 
near $\Sigma$, with the free data consisting of just the
3-metric there. The existence and uniqueness results obtained are used
to prove the Vanishing Weyl Tensor conjecture described in the
Introduction.

\subsection{Conformal gauge fixing}
Suppose then that $\Omega$ is the conformal factor relating the physical
metric $\tilde{g}_{ab}$ to the unphysical metric $g_{ab}$, by means
of
\begin{equation}\tilde{g}_{ab}=\Omega^{2}g_{ab}\end{equation}
with $\Omega=0$ at $\Sigma$ in $M$. Suppose also that the
stress-energy-momentum tensor in~$\tilde{M}$~is that appropriate to a
polytropic perfect fluid:
\begin{equation}\tilde{T}_{ab}=(\rho + p)\tilde{u}_{a}\tilde{u}_{b} -
p\tilde{g}_{ab}\end{equation}
\begin{equation}p= (\gamma - 1) \rho~~~~1\leq\gamma \leq
2\end{equation}
To make analytical progress we assume that the rescaled fluid flow given by~$u_{a}=\Omega^{-1}\tilde{u}_{a}$~remains regular and non-zero near~$\Sigma$.

\noindent From the work of Newman we know that if one demands that
$\Omega$ be smooth with non-zero gradient at $\Sigma$ then the Einstein
equations imply that the
value of the polytropic index must be $\gamma=4/3$, the value for
radiation. So if one wants to consider other equations of state then a
different differentiability conditon on $\Omega$ is required. To see
what this should be, consider the FRW cosmologies, where one may take the scale factor $R(t)$
for $\Omega$. With a perfect fluid source, the field equations imply
that the behaviour of $dR$ at $R=0$ is tied to the value of
the polytropic index $\gamma$ according to
\begin{equation}dR\sim~\tau^{\frac{4-3\gamma}{3\gamma-2}}~\textrm{as}~\tau\rightarrow~0\end{equation}
in terms of conformal time $\tau$ which is a smooth coordinate in $M$.
In view of this, we will impose the following condition:
\begin{equation}\textrm{At}~\Sigma~,~
Z\equiv\Omega^{~(3\gamma-2)/2}~~\textrm{is smooth, and}~\nabla_{a}Z\neq~0\end{equation}
It is a result of Goode (1987) that the physical velocity field ~$\tilde{u}_{a}$~ must be
hypersurface orthogonal, and this will also be true for the unphysical
velocity field ~$u_{a}\equiv\Omega^{-1}\tilde{u}_{a}$~. We know also
that~$u_{a}$~meets~$\Sigma$~orthogonally. Thus near~$\Sigma$~in ~$M$
\begin{equation}u_{a}=\frac{1}{V}\hat{Z}_{,a}~~~~~~~~V^{2}=g^{ab}\hat{Z}_{,a}\hat{Z}_{,b}\end{equation}
for some smooth cosmic time function $\hat{Z}$ on~$M$, with~$\hat{Z}=0$~on~$\Sigma$. A natural choice of
conformal factor is then given by the following lemma:
\\

\textbf{Lemma 3.1}~If (20) holds then one may take the conformal
factor $\Omega$ to be a function of the cosmic time defined by the
fluid velocity.

\textit{Proof}~We know~$u_{a}=\psi\nabla_{a}\hat{Z}$~for some
smooth~$\psi,~\hat{Z}$~in~$M$~near~$\Sigma$, with~$\hat{Z}=0$~at~$\Sigma$.
Now~$\tilde{g}_{ab}=\hat{\Omega}^{2}~\hat{g}_{ab}$~,~where~$\hat{g}_{ab}=(\Omega^{2}/\hat{\Omega}^{2})g_{ab}$,~and\linebreak[4]
$\hat{\Omega}=(\hat{Z})^{(1/q)}~(q=(3\gamma-2)/2)$~.
So~$\hat{g}_{ab}=(Z/\hat{Z})^{(2/q)}g_{ab}$~, where~$Z$~ is as in
(20). Since~$\nabla_{a}Z,~\nabla_{a}\hat{Z}$~are non-zero in~$M$~,
one has~$Z=\hat{Z}~f$~, for some smooth function~$f$~in~$M$, which
does not vanish at~$\Sigma$.~Therefore~$f^{(2/q)}$~is smooth
at~$\Sigma$, and~$\hat{g}_{ab}$~is smooth in~$M$~$\square$
\subsection{Conformal field equations in~$M$}
We now make explicit the field equations we wish to solve
for~$g_{ab}$~in~$M$.

If we make the conformal gauge choice of Lemma 3.1, with~$Z$~as in
(20) and~$u_{a},~V$~as in the unhatted version of
(21), then for a physical stress
tensor of the form (17) the conformally transformed Einstein-perfect fluid equations
for~$g_{ab}$~in~$M$~are (Tod 1991):
\begin{eqnarray}R_{ab}-\frac{4}{3\gamma-2}\frac{1}{Z}\nabla_{a}\nabla_{b}Z+\frac{12\gamma}{(3\gamma-2)^{2}}\frac{1}{Z^{2}}Z_{a}Z_{b}(1-G)\nonumber\\
-g_{ab}\left(\frac{2}{(3\gamma-2)}\frac{1}{Z}\square Z+\frac{6(2-\gamma)}{(3\gamma-2)^{2}}\frac{V^{2}}{Z^{2}}(1-G)\right)=0\end{eqnarray}
Here ~$Z_{a}=\nabla_{a}Z$~and~$G$~parametrizes the density according to
\begin{equation}\rho=\frac{12}{(3\gamma-2)^{2}}\frac{V^{2}G}{Z^{6\gamma/(3\gamma-2)}}\end{equation}

This is the field equation in~$M$~for which we wish to formulate the
Cauchy problem.

Firstly there is information in the contracted Bianchi identity
applied to (22). Projecting along and perpendicular to~$u_{a}$~gives,
respectively
\begin{equation}\square Z+\frac{(2-\gamma)}{\gamma}\frac{Z^{b}V_{b}}{V}+\frac{1}{\gamma}\frac{Z^{b}G_{b}}{G}=0\end{equation}
\begin{equation}(2-\gamma)\frac{V_{a}}{V}-(\gamma-1)\frac{G_{a}}{G}+\left((\gamma-1)\frac{Z^{c}G_{c}}{G}-(2-\gamma)\frac{Z^{c}V_{c}}{V}\right)\frac{Z_{a}}{V^{2}}=0\end{equation}
where~$V_{a}=\nabla_{a}V,~G_{a}=\nabla_{a}G$.

If~$\gamma\neq 1$~then (25) can be solved to give
\begin{equation}G=V^{P}g(Z)~~~~~P=(2-\gamma)/(\gamma-1)\end{equation}
for some function~$g$.

The case~$\gamma=1$~will be treated shortly. Now we aim to show that~$g$~can
be set to~$1$~by a change of cosmic time~$Z$. Multiplying (22)
by~$Z^{2}$~and letting~$Z\rightarrow 0$~we get~$G\rightarrow 1$. Also,
we see that~~$Z^{-1}(1-G)$~~has a limit as~$Z\rightarrow 0$. Now
multiply (22) by~$ZZ^{b}$~and let~$Z\rightarrow 0$~to find
that~$V_{b}$~is proportional to~$Z_{b}$~in the limit. This means
that~$V$~must be constant on~$\Sigma$~and we can take this constant to
be~$1$~wlog. From (26) this entails~$g\rightarrow
1$~as~$Z\rightarrow 0$.

Suppose now that we do (16) two different ways:
\begin{equation}\tilde{g}_{ab}=\Omega^{2}g_{ab}=\hat{\Omega}^{2}\hat{g}_{ab}\end{equation}
then from (23) since these both correspond to the same~$\rho$,
\begin{equation}V^{2}G(\hat{Z})^{6\gamma/(3\gamma-2)}=\hat{V}^{2}\hat{G}Z^{6\gamma/(3\gamma-2)}\end{equation}
Now use (26) to get
\begin{equation}\left(\frac{g}{\hat{g}}\right)^{(\gamma-1)/\gamma}=\frac{\hat{V}}{V}\left(\frac{Z}{\hat{Z}}\right)^{6(\gamma-1)/(3\gamma-2)}\end{equation}
Finally~$\hat{V}/{V}=(\hat{\Omega}/\Omega)(d\hat{Z}/dZ)$~so that
\begin{equation}\frac{d\hat{Z}}{dZ}=\left(\frac{g}{\hat{g}}\right)^{(\gamma-1)/\gamma}\left(\frac{\hat{Z}}{Z}\right)^{2(3\gamma-4)/(3\gamma-2)}\end{equation}

Solving (30) for~$\hat{Z}$~with~$\hat{g}=1$~determines a new choice
of unphysical spacetime in which~$g=1$~, and (26) becomes
\begin{equation}G=V^{(2-\gamma)/(\gamma-1)}\end{equation}
while (24) becomes
\begin{equation}\square Z+~P\frac{Z^{b}V_{b}}{V}=0\end{equation}
If we let~$h_{ab}=g_{ab}-u_{a}u_{b}$~ be the metric on the surfaces of
constant~$Z$~, and~$K_{ab}=\frac{1}{2}\mathcal{L}_{u}h_{ab}$~the
second fundamental form, then we have the identity
\begin{equation}\square Z=V\left(K+\frac{\partial
V}{\partial Z}\right)\end{equation}
where~$K=h^{ab}K_{ab}$.\\
Therefore (32) can be written as
\begin{equation}\frac{\partial V}{\partial Z}=~(1-\gamma)K\end{equation}

The field equations to be solved are now (22), (31) and (34)

Returning to the case~$\gamma=1$~we find from (25) that~$V$~is a
function only of~$Z$. Now we can use a change of cosmic time to
set~$V=1$~. To see this note that if~$Z, \hat{Z}$~are two choices of
cosmic time with corresponding functions~$\Omega, V$~and~$\hat{\Omega},
\hat{V}$~then
\begin{equation}\frac{d\hat{Z}}{dZ}=\frac{\hat{V}\hat{\Omega}}{V\Omega}=\frac{\hat{V}}{V}\left(\frac{\hat{Z}}{Z}\right)^{2}\end{equation}
\\From (22) we again deduce that~$V\rightarrow$~constant
on~$\Sigma$~and we can take this constant to be~$1$. Now we may
put~$\hat{V}=1$~in (30) and solve for~$\hat{Z}$~to get the required
result.

In this case (24) becomes
\begin{equation}\square Z+\frac{Z^{b}G_{b}}{G}=0\end{equation}
and thus by (33) one gets
\begin{equation}\frac{\partial G}{\partial Z}=-GK\end{equation}
and the field equations are (22) and (37)

\subsection{Solving `reduced' field equations for~$1<\gamma<2$}

If one wishes to give data for (22), (34) at~$\Sigma$~then the
equations impose constraints. We've seen that~$V$~must be equal to
one, and one also gets~$K_{ab}=0$~from (22) (or see (41) below). 
The equations impose no
constraints on the initial 3-metric~$h_{ab}$~, and this is all the
free data at~$\Sigma$.

The aim now is to cast the equations (22), (34) in a symmetric hyperbolic
form for which Theorem 2 applies.

We will use coordinates~$Z,~x^{i}$~, with the~$x^{i}$~being carried
along the integral curves of~~$\nabla^{a}Z$, on a manifold of the
form~$\Sigma~\mathbf{\times}~[0,T]$.

In these coordinates the metric~$g$~assumes the form
\begin{equation}ds^{2}=\frac{1}{V^{2}}dZ^{2}~+~h_{ij}dx^{i}dx^{j}\end{equation}
with ~$h_{ij}$~negative definite.
\\Making a~$3+1$~split of equations (22) gives
\begin{equation}R_{zz}=\frac{2K}{VZ}(4-3\gamma)/(3\gamma-2)~+~\frac{6}{Z^{2}}(V^{P}-1)/(3\gamma-2)\end{equation}
\begin{equation}R_{zi}=\frac{4}{Z}\frac{1}{(3\gamma-2)}\partial_{i}\log
V\end{equation}
\begin{equation}R_{ij}=\frac{4}{(3\gamma-2)}\frac{V}{Z}K_{ij}~+~h_{ij}\left(\frac{2V(2-\gamma)}{(3\gamma-2)}\frac{K}{Z}~+~\frac{6(2-\gamma)}{(3\gamma-2)^{2}}\frac{V^{2}}{Z^{2}}(1-V^{P})\right)\end{equation}

The~$\frac{1}{Z^{2}}$~terms which appear here are undesirable with
regard to application of the Newman-Claudel theorem. We therefore
introduce a new variable~$\zeta$~ according to
\begin{equation}\zeta=\frac{6(2-\gamma)}{(3\gamma-2)^{2}}\frac{(V^{P}-1)}{Z}\end{equation}
\\From (34) there follows
\begin{equation}\frac{\partial\zeta}{\partial
Z}=-\frac{1}{Z}\zeta~-~\frac{1}{Z}\frac{6(2-\gamma)^{2}}{(3\gamma-2)^{2}}V^{(P-1)}K\end{equation}

By definition one has
\begin{equation}R_{ab}=-\frac{1}{2}g^{cd}\partial^{2}_{cd}g_{ab}~+~\partial_{(a}\Gamma_{b)}~+~H_{ab}\end{equation}
where
\begin{equation}H_{ab}=2g^{ef}\left(g_{c(a}\Gamma^{d}_{b)f}\Gamma^{c}_{de}~+~g_{cd}\Gamma^{d}_{f[b}\Gamma^{c}_{e]a}\right)~~,~~\Gamma_{a}=g_{ad}g^{bc}\Gamma^{d}_{bc}\end{equation}
(Note that $H_{ab}$ is not the magnetic part of the Weyl tensor.)
For a metric of the form (38) one therefore has
\begin{equation}\Gamma_{i}=\partial_{i}\log V~+~h^{mn}(\partial_{m}h_{ni}-\frac{1}{2}\partial_{i}h_{mn})\end{equation}
\begin{equation}\Gamma_{z}=-\partial_{z}\log V~+~\frac{1}{2}h_{ij}k^{ij}\end{equation}
where~$k^{ij}=\partial_{z}h^{ij}~(\Rightarrow~K_{ij}=-\frac{1}{2}Vh_{im}h_{jn}k^{mn}~,~K=-\frac{1}{2}V(h_{ij}k^{ij})~)$.

\noindent Also
\begin{displaymath}H_{ij}=2h^{kl}\left(h_{m(i}\Gamma^{n}_{j)l}\Gamma^{m}_{nk}~+~h_{mn}\Gamma^{n}_{l[j}\Gamma^{m}_{k]i}\right)\end{displaymath}\begin{displaymath}-~(\partial_{i}\log
V)(\partial_{j}\log V)~-\Gamma^{k}_{ij}\partial_{k}\log
V~+~\frac{1}{2}V^{2}(\partial_{z}\log
V)h_{ki}h_{jl}k^{kl}\end{displaymath}\begin{equation}+~\frac{1}{2}V^{2}h_{ki}h_{jl}h_{mn}k^{mk}k^{ln}~-\frac{1}{4}V^{2}h_{ki}h_{jl}h_{mn}k^{kl}k^{mn}\end{equation}\begin{displaymath}H_{iz}=(h_{k[j}\partial_{i]}\log
V)k^{jk}\end{displaymath}\begin{equation}+\frac{1}{2}h_{ij}h_{mn}(h^{pq}k^{mj}-h^{mj}k^{pq})\Gamma^{n}_{pq}\end{equation}\begin{displaymath}H_{zz}=~2(\partial_{z}\log
V)^{2}~+~\frac{1}{2}(\partial_{z}\log V)h_{mn}k^{mn}~-~\frac{1}{4}h_{ij}h_{kl}k^{ik}k^{jl}\end{displaymath}\begin{equation}+\frac{1}{V^{2}}h^{ij}\left(
(\partial_{i}\log V)(\partial_{j}\log
V)-\Gamma^{k}_{ij}\partial_{k}\log V\right)\end{equation}
\\
Define~$\gamma_{ijk}\equiv\partial_{k}h_{ij}$~. Then from the
definitions (46), (47) one gets
\begin{equation}\partial_{z}\Gamma_{i}=~\partial_{i}\Gamma_{z}-h_{ik}\partial_{j}k^{jk}+2\partial_{i}\partial_{z}\log
V-h_{ik}h^{mn}k^{jk}\gamma_{mjn}\end{equation}
Now (34), (47) imply
\begin{equation}\Gamma_{z}=\frac{1}{2}(2-\gamma)h_{ij}k^{ij}~=~P\partial_{z}\log
V\end{equation}
whereby (51) becomes
\begin{equation}\partial_{z}\Gamma_{i}=\left(\frac{\gamma}{2-\gamma}\right)\partial_{i}\Gamma_{z}-h_{ik}\partial_{j}k^{jk}-h_{ik}h^{mn}k^{jk}\gamma_{mjn}\end{equation}
We now choose new independent variables in such a way that the field
equations for these quantities become symmetric hyperbolic.\\
Define the positive constants~$E, F, G$~by
\begin{equation}E^{2}=\frac{24\gamma(2-\gamma)}{(\gamma-1)^{2}}~~,~~F^{2}=\frac{12\gamma}{(\gamma-1)}~~,~~G^{2}=\frac{4\gamma}{(2-\gamma)}\end{equation}
and define the following new variables
\begin{equation}\chi_{i}\equiv~E\left(\partial_{i}\log V+\frac{1}{2P}\Gamma_{i}\right)~~,~~\nu\equiv~F\Gamma_{z}~~,~~\xi_{i}\equiv~G\Gamma_{i}\end{equation}
\\
In terms of the new variables the field equations can be written as follows:
\begin{equation}\partial_{z}h^{ij}=k^{ij}\end{equation}
\begin{equation}\partial_{z}h_{ij}=-h_{im}h_{jn}k^{mn}\end{equation}
\begin{equation}\partial_{z}V=\frac{1}{2}(\gamma-1)V\left(\frac{h_{(mn)}k^{mn}}{(1-5\gamma)}-\frac{10\gamma}{(2-\gamma)(1-5\gamma)}\frac{\nu}{F}\right)\end{equation}
\begin{displaymath}\partial_{z}\zeta=\frac{(2-\gamma)}{2(1-5\gamma)}\left(h_{(mn)}k^{mn}-\frac{10\gamma}{(2-\gamma)}\frac{\nu}{F}\right)\end{displaymath}
\begin{equation}-\frac{1}{Z}\left(\zeta-\frac{3(2-\gamma)^{2}}{(3\gamma-2)^{2}(1-5\gamma)}\left(h_{(mn)}k^{mn}-\frac{10\gamma}{(2-\gamma)}\frac{\nu}{F}\right)\right)\end{equation}
\begin{equation}-h^{(im)}h^{(jn)}h^{(kl)}\partial_{z}\gamma_{mnl}=~h^{(kl)}\partial_{l}k^{(ij)}~+~h^{kl}(h^{im}k^{jn}+h^{jn}k^{im})\gamma_{mnl}\end{equation}
\begin{displaymath}-\left(\frac{3\gamma-2}{2\gamma}\right)h^{(ij)}\partial_{z}\xi_{i}=~\frac{3}{2}\frac{G}{F}h^{(ij)}\partial_{i}\nu~+~G\left(\frac{\gamma-2}{2\gamma}\right)\partial_{k}k^{(jk)}\end{displaymath}\\\begin{displaymath}+~G\left(\frac{\gamma-2}{2\gamma}\right)h^{mn}k^{lj}\gamma_{mln}~+~2Gh^{ij}\hat{H}_{iz}
\end{displaymath}\begin{equation}-~\frac{8G}{Z(3\gamma-2)}h^{(ij)}\left(\frac{\chi_{i}}{E}-\frac{1}{2PG}\xi_{i}\right)\end{equation}
\begin{displaymath}V^{2}h_{(im)}h_{(jn)}\partial_{z}k^{mn}=~h^{(mn)}\partial_{m}\gamma_{(ij)n}-~\frac{2}{G}\partial_{(i}\xi_{j)}\end{displaymath}
\begin{displaymath}+~2V^{2}k^{km}k^{pr}h_{(im)}h_{(jp)}h_{(kr)}-~2\hat{H}_{ij}\end{displaymath}
\begin{equation}+~\frac{2V^{2}}{Z}\left(-\frac{2}{(3\gamma-2)}h_{(im)}h_{(jn)}k^{mn}-h_{(ij)}\left(\zeta-\frac{h_{(mn)}k^{mn}}{(3\gamma-2)}+\frac{2(\gamma-3)}{(3\gamma-2)(\gamma-2)}\frac{\nu}{F}\right)\right)\end{equation}
\begin{displaymath}\left(\frac{V^{2}}{2-\gamma}\right)\partial_{z}\nu=~-\frac{F}{E}h^{(ij)}\partial_{i}\chi_{j}+~\frac{F}{2PG}h^{(ij)}\partial_{i}\xi_{j}\end{displaymath}
\begin{displaymath}+~2Fh^{ij}\left(\frac{\chi_{i}}{E}-\frac{\xi_{i}}{2PG}\right)\left(\frac{\chi_{j}}{E}-\frac{\xi_{j}}{2PG}\right)~+~\frac{2V^{2}}{FP^{2}}\nu^{2}~-~V^{2}F\hat{H}_{zz}\end{displaymath}
\begin{equation}+\frac{V^{2}F}{Z}\left(\left(\frac{3\gamma-2}{2-\gamma}\right)\zeta+\frac{h_{(mn)}k^{mn}}{(3\gamma-2)}+\frac{2(3\gamma-5)}{(2-\gamma)(3\gamma-2)}\frac{\nu}{F}\right)\end{equation}
\begin{displaymath}-h^{(ij)}\partial_{z}\chi_{i}=~-\frac{E}{2PF}h^{(ij)}\partial_{i}\nu~+~\frac{E}{P}h^{(ij)}\hat{H}_{zi}\end{displaymath}
\begin{equation}+~\frac{4E}{(3\gamma-2)PZ}h^{(ij)}\left(\frac{\xi_{i}}{2PG}-\frac{\chi_{i}}{E}\right)\end{equation}
\\
where
\begin{displaymath}\hat{H}_{zz}=~-\frac{1}{4}h_{ij}h_{kl}k^{ik}k^{jl}~+~\frac{1}{V^{2}}h^{ij}\left(\frac{\chi_{i}}{E}-\frac{\xi_{i}}{2PG}\right)\left(\frac{\chi_{j}}{E}-\frac{\xi_{j}}{2PG}\right)\end{displaymath}
\begin{equation}-h^{ij}\hat{\Gamma}^{k}_{ij}\left(\frac{\chi_{k}}{E}-\frac{\xi_{k}}{2PG}\right)\end{equation}
\begin{displaymath}\hat{H}_{iz}=~h_{k[j}\left(\frac{1}{E}\chi_{i]}-\frac{1}{2PG}\xi{i]}\right)k^{jk}\end{displaymath}
\begin{equation}+~\frac{1}{2}h_{ij}h_{mn}(h^{pq}k^{mj}-h^{mj}k^{pq})\hat{\Gamma}^{n}_{pq}\end{equation}
\begin{displaymath}\hat{H}_{ij}=~2h^{(kl)}\left(h_{m(i}\hat{\Gamma}^{n}_{j)l}\hat{\Gamma}^{m}_{nk}+h_{mn}\hat{\Gamma}^{n}_{l[j}\hat{\Gamma}^{m}_{k]i}\right)\end{displaymath}\begin{displaymath}-\left(\frac{\chi_{i}}{E}-\frac{\xi{i}}{2PG}\right)\left(\frac{\chi_{j}}{E}-\frac{\xi_{j}}{2PG}\right)-~\hat{\Gamma}^{k}_{ij}\left(\frac{\chi_{k}}{E}-\frac{\xi_{k}}{2PG}\right)\end{displaymath}
\begin{equation}+~\frac{1}{2}V^{2}h_{k(i}h_{j)l}k^{km}k^{ln}h_{mn}+~\frac{1}{4}V^{2}(\gamma-2)h_{mn}k^{mn}h_{k(i}h_{j)l}k^{kl}\end{equation}
and
\begin{equation}\hat{\Gamma}^{k}_{ij}=\bar{\Gamma}^{k}_{(ij)}~;~\bar{\Gamma}^{k}_{ij}=\frac{1}{2}h^{kl}(\gamma_{lij}+\gamma_{jli}-\gamma_{ijl})\end{equation}

In (58), (59), (62), (63) we have used the first of (52) to
partially substitute for~$h_{ij}k^{ij}$~, which will be useful later
on. Equation (62) comes from (44) and (41). Equation (63)
comes from (44), (39), and the second of (52). Equation (61)
comes from (40), (44), and (53) was used to partially substitute
for~$\partial_{i}\Gamma_{z}$. Equation (64) comes from (44),
(40), and the second of (52)\\
We now seek a solution of  the `reduced' equations (56)-(68),\linebreak
 where~$\gamma_{ijk},~\chi_{i},~\nu,~\xi_{i}$~ are to be treated as
independent variables, with no reference to their original definitions. We will suppose that the
coordinates~$x^{i}$~are chosen to satisfy the harmonic gauge condition
\begin{equation}h^{ij}\Gamma^{k}_{ij}=0\end{equation}
on~$\Sigma$.

\noindent The initial data for the reduced equations are as follows:\\
$h_{ij}$ is free data at~$\Sigma$~, and is chosen negative definite and
symmetric.~$h^{ij}$~is chosen so that
~$h^{ij}h_{jk}=\delta^{i}_{k}$, and choose also~$\gamma_{ijk}=\partial_{k}h_{ij}$. Since the second fundamental form
vanishes at~$\Sigma$~one must have~$k^{ij}=0$~there, and one also
has~$V=1$. From the first of (52) and (53), (56), (55), (69) there follows~$\nu=~\zeta=~\chi_{i}=~\xi_{i}=0$

It is possible, using the formalism of Newman (1993b), to write
(56)-(68) explicitly in the following form
\begin{equation}A^{0}(u)\partial_{z}u=~A^{i}(u)\partial_{i}(u)+~B(u)u+~\frac{1}{Z}C(u)u\end{equation}
where~$u$~stands for the field variables written as a
vector,~$A^{0}$~is positive definite and symmetric, the~$A^{i}$~are
symmetric, and~$A^{\alpha}, B, C$~are polynomial in their
arguments (see Appendix for details). One has~$C(u)(u(0))=0$~ for all~$u$, and the eigenvalues
of~$((A^{0})^{-1}C)(u(0))$~satisfy  one of the following
\begin{displaymath}\lambda=0,~~\lambda=\frac{4(\gamma-2)}{P(3\gamma-2)^{2}}\end{displaymath}
\begin{displaymath}\lambda^{3}+~\frac{3(2-\gamma)}{(3\gamma-2)}\lambda^{2}+~\frac{12(2-\gamma)(5\gamma-3)}{(3\gamma-2)^{2}(5\gamma-1))}\lambda+~\frac{32(13\gamma-8)}{(5\gamma-1)(3\gamma-2)^{3}}=0\end{displaymath}
\\
Hence for~$1<\gamma<2$,~no~$\lambda$~is positive. It follows that
(56)-(68)~satisfy all the requirements of Theorem 2, and thus there exists a solution~$u$~unique in a suitable
differentiability class. Differentiability is discussed at the end of
section 4.4.\\
One now wishes to know whether the metric~$g$~obtained from ~$u$~via
(38) is actually a solution of the conformal Einstein equations
(22). i.e. whether the definitions of
~$\gamma_{ijk},~\chi_{i},~\nu,~\xi_{i}$~ can be recovered from
(56)-(58)~. The answer is yes, as is shown in the next section.

\subsection{Recovering the conformal Einstein equations}
We now show that if~$u$~is the solution of (56)-(58) as
above, then as long as~$h_{(ij)}$~remains negative
definite,~$h_{ij}$~is in fact  symmetric, and~$g$~given by (38) is a solution
of the conformal Einstein-perfect fluid equations (22). The key step
will be, following Newman (1993b), to use the contracted Bianchi identities to
show that~$\nu/F$~and~$\xi_{i}/G$~agree with the definitions
(46)-(47) of~$\Gamma_{z}$~and~$\Gamma_{i}$.\\
\\Suppose then that~$h_{ij}$~is negative definite. Equation (62) then
gives
\begin{displaymath}\partial_{z}(Z^{4/(3\gamma-2)}k^{[ij]})=0\end{displaymath}
Hence~$k^{[ij]}=0$~for all $Z$. Then
(56)-(57) give $h_{[ij]}=h^{[ij]}=0$~for all $Z$. One also gets
\begin{displaymath}\partial_{z}(h^{kj}h_{ji}-\delta^{k}_{i})=-h_{im}(h^{kj}h_{jn}-\delta^{k}_{n})k^{mn}\end{displaymath}
A Gronwall estimate (Racke 1992) therefore
gives~$h^{kj}h_{ji}=\delta^{k}_{i}$~for all $Z$. Now (56), (60)
imply
\begin{displaymath}\partial_{z}(\gamma_{ijk}-\partial_{k}h_{ij})=h_{jm}k^{mn}(\gamma_{ink}-\partial_{k}h_{in})+h_{im}k^{mn}(\gamma_{njk}-\partial_{k}h_{nj})\end{displaymath}
So a Gronwall estimate
implies~$\gamma_{ijk}=\partial_{k}h_{ij}$~for all $Z$.\\
Define ~$\hat{\zeta}$~by
\begin{displaymath}\hat{\zeta}=\frac{6(2-\gamma)}{(3\gamma-2)^{2}}\frac{(V^{P}-1)}{Z}\end{displaymath}
then (58), (59) give
\begin{displaymath}\partial_{z}(Z(\zeta-\hat{\zeta}))=0\end{displaymath}
and hence the definition (52) is recovered.\\
Now
write~$\hat{\Gamma}_{z}=\frac{\nu}{F}~,~\hat{\Gamma}_{i}=\frac{\xi_{i}}{G}~,~\psi=\frac{\chi_{i}}{E}$.
Then (61), (64) imply
\begin{displaymath}\partial_{z}\left(\left(\frac{2-3\gamma}{2\gamma}\right)\hat{\Gamma}_{i}+2P\psi_{i}+\left(\frac{\gamma-2}{2\gamma}\right)\Gamma_{i}+\frac{2(\gamma-2)}{(\gamma-1)}\partial_{i}\log
V\right)=\end{displaymath}
\begin{displaymath}\partial_{i}\left(\frac{5}{2}\hat{\Gamma}_{z}+\frac{(\gamma-2)(5\gamma-1)}{2\gamma(\gamma-1)}\partial_{z}\log
V+\left(\frac{\gamma-2}{4\gamma}\right)h_{ij}k^{ij}\right)\end{displaymath}
and now (58) gives
\begin{equation}\left(\frac{2-3\gamma}{2\gamma}\right)\hat{\Gamma}_{i}+2P\psi_{i}+\left(\frac{\gamma-2}{2\gamma}\right)\Gamma_{i}-2P\partial_{i}\log
V=0\end{equation}
Note that once we have established~~$\hat{\Gamma}_{i}=\Gamma_{i}$~~then
(71) will
imply\linebreak[4] $\psi_{i}=\partial_{i}\log V+~\frac{1}{2P}\Gamma_{i}$~as desired.\\
\\From the definition of~$\Gamma_{z}$~and (58) there follows
\begin{equation}\partial_{z}\log V=\left(\frac{\gamma-1}{2-\gamma}\right)\hat{\Gamma}_{z}+\frac{(1-\gamma)}{2(1-3\gamma)}(\hat{\Gamma}_{z}-\Gamma_{z})\end{equation}
while (71) implies
\begin{equation}\psi_{i}-\frac{1}{2P}\hat{\Gamma}_{i}=\frac{(\gamma-1)}{4\gamma}(\Gamma_{i}-\hat{\Gamma}_{i})+\partial_{i}\log
V\end{equation}
and (64) implies
\begin{equation}-\partial_{z}\psi_{i}=-\frac{1}{2P}\partial_{i}\hat{\Gamma}_{z}+\frac{1}{P}\hat{H}_{zi}+\frac{(4P(3\gamma-2))^{-1}}{Z}\left(\frac{1}{2P}\hat{\Gamma}_{i}-\psi_{i}\right)\end{equation}
Equations (72)-(74) and (51) now combine to yield
\begin{displaymath}-2\hat{H}_{zi}+\frac{8}{Z(3\gamma-2)}\left(\psi_{i}-\frac{1}{2P}\hat{\Gamma}_{i}\right)=-h_{ik}\partial_{j}k^{jk}-h_{ik}h^{mn}k^{jk}\gamma_{mjn}\end{displaymath}
\begin{equation}+\left(\frac{2-3\gamma}{1-3\gamma}\right)\partial_{i}(\Gamma_{z}-\hat{\Gamma}_{z})+\frac{2}{(2-\gamma)}\partial_{i}\hat{\Gamma}_{z}+\frac{(3\gamma-2)}{2\gamma}\partial_{z}(\hat{\Gamma}_{i}-\Gamma_{i})\end{equation}
\\
Define a symmetric tensor~$S_{ab}$~by
\begin{displaymath}S_{ij}=\frac{V^{2}}{Z}\left\{-\frac{2}{(3\gamma-2)}h_{im}h_{jn}k^{mn}\right.\end{displaymath}
\begin{equation}\left.-h_{ij}\left(\zeta+\frac{(2-\gamma)}{(3\gamma-2)}\left(\frac{1}{(\gamma-2)}h_{mn}k^{mn}+\frac{2(3-\gamma)}{(\gamma-2)^{2}}\hat{\Gamma}_{z}\right)\right)\right\}\end{equation}
\begin{equation}S_{iz}=\frac{4}{Z}\frac{1}{(3\gamma-2)}\left(\psi_{i}-\frac{1}{2P}\hat{\Gamma}_{i}\right)\end{equation}
\begin{equation}S_{zz}=\frac{1}{Z}\left\{\left(\frac{(3\gamma-2)}{2-\gamma}\right)\zeta+\frac{1}{3\gamma-2}h_{mn}k^{mn}+\frac{2(3\gamma-5)}{(2-\gamma)(3\gamma-2)}\hat{\Gamma}_{z}\right\}\end{equation}
\\
Also define
~$T_{ab}=S_{ab}-\frac{1}{2}Sg_{ab}$~, where~$S=g^{ab}S_{ab}$.

Then
\begin{equation}T_{ij}=\frac{V^{2}}{Z}\left\{-\frac{2}{(3\gamma-2)}h_{im}h_{jn}k^{mn}-h_{ij}\left(\frac{2\zeta}{P}-\frac{4}{(3\gamma-2)}\hat{\Gamma}_{z}\right)\right\}\end{equation}
\begin{equation}T_{iz}=\frac{4}{Z}\frac{1}{(3\gamma-2)}\left(\psi_{i}-\frac{1}{2P}\hat{\Gamma}_{i}\right)\end{equation}
\begin{equation}T_{zz}=\frac{1}{Z}\left\{\frac{2}{(2-\gamma)}+\frac{4}{(2-\gamma)(3\gamma-2)}\hat{\Gamma}_{z}\right\}\end{equation}
Equations (62), (64), (63) can now be written, respectively, as
\begin{equation}S_{ij}=R_{ij}+\partial_{(i}(\hat{\Gamma}_{j)}-\Gamma_{j)})+(\hat{H}_{ij}-H_{ij})\end{equation}
\begin{displaymath}S_{iz}=R_{iz}+(\hat{H}_{iz}-H_{iz})\end{displaymath}\begin{equation}+\frac{1}{2}\left(\frac{1+2\gamma}{3\gamma-1}\right)\partial_{i}(\hat{\Gamma}_{z}-\Gamma_{z})+\left(\frac{3\gamma-2}{4\gamma}\right)\partial_{z}(\hat{\Gamma}_{i}-\Gamma_{i})\end{equation}
\begin{displaymath}S_{zz}=R_{zz}+(\hat{H}_{zz}-H_{zz})+\frac{(5\gamma-1)}{2(3\gamma-1)}\partial_{z}(\hat{\Gamma}_{z}-\Gamma_{z})\end{displaymath}
\begin{displaymath}+\left(\frac{1-\gamma}{4\gamma
V^{2}}\right)h^{ij}\partial_{i}(\hat{\Gamma}_{j}-\Gamma_{j})-\frac{2}{V^{2}}\left(\frac{\gamma-1}{4\gamma}\right)^{2}h^{ij}(\hat{\Gamma}_{i}-\Gamma_{i})(\hat{\Gamma}_{j}-\Gamma_{j})\end{displaymath}
\begin{equation}+\left(\frac{1-\gamma}{V^{2}\gamma}\right)h^{ij}(\Gamma_{i}-\hat{\Gamma}_{i})\partial_{j}\log
V+\frac{1}{2}\left(\frac{1-\gamma}{1-3\gamma}\right)^{2}(\hat{\Gamma}_{z}-\Gamma_{z})^{2}+\frac{2(\gamma-1)^{2}}{(2-\gamma)(1-3\gamma)}\hat{\Gamma}_{z}(\Gamma_{z}-\hat{\Gamma}_{z})\end{equation}

If (56)-(68) hold, then one calculates the following
\begin{displaymath}\nabla_{a}T^{a}_{i}=\frac{2}{(3\gamma-2)}\frac{V^{2}}{Z}\Bigg\{\left(\frac{4\gamma-3}{1-3\gamma}\right)\partial_{i}(\Gamma_{z}-\hat{\Gamma}_{z})+\left(\frac{4\gamma-3}{2\gamma}\right)\partial_{z}(\Gamma_{i}-\hat{\Gamma}_{i})\end{displaymath}
\begin{displaymath}+\left(\frac{7\gamma-1}{3\gamma-1}\right)\partial_{i}\log
V(\hat{\Gamma}_{z}-\Gamma_{z})+\left(\frac{\gamma-1}{4\gamma}\right)(\Gamma_{j}-\hat{\Gamma}_{j})h_{im}k^{mj}\end{displaymath}
\begin{equation}+\frac{3}{Z}\frac{(2-\gamma)(\gamma-1)}{2(3\gamma-2)}(\Gamma_{i}-\hat{\Gamma}_{i})+\left(\frac{\gamma-1}{4\gamma}\right)\left(\left(\frac{1+7\gamma}{1-3\gamma}\right)\Gamma_{z}+\frac{(1-\gamma)(8-\gamma)}{(2-\gamma)(1-3\gamma)}\hat{\Gamma}_{z}\right)(\Gamma_{i}-\hat{\Gamma}_{i})\Bigg\}\end{equation}
\\
\begin{displaymath}\nabla_{a}T^{a}_{z}=\frac{V^{2}}{Z}\left\{\left(\frac{\hat{\Gamma}_{z}-\Gamma_{z}}{Z}\right)\left(\frac{2(13\gamma-8)-6(V^{P}-1)(5\gamma^{2}-4\gamma+3)}{(1-3\gamma)(3\gamma-2)^{2}}\right)\right.\end{displaymath}
\begin{displaymath}\left.+2\hat{\Gamma}_{z}(\hat{\Gamma}_{z}-\Gamma_{z})\frac{(1-\gamma)(5\gamma-4)}{(3\gamma-2)(2-\gamma)(1-3\gamma)}\right\}\end{displaymath}
\begin{equation}+\frac{1}{Z}\left\{\frac{(1-\gamma)}{(3\gamma-2)\gamma}h^{kj}(\partial_{j}\log
V)(\hat{\Gamma}_{k}-\Gamma_{k})+\frac{4}{(3\gamma-2)}\left(\frac{\gamma-1}{4\gamma}\right)^{2}h^{ij}(\hat{\Gamma}_{i}-\Gamma_{i})(\hat{\Gamma}_{j}-\Gamma_{j})\right\}\end{equation}
\\
Now write
~$\Delta_{ab}=S_{ab}-R_{ab}=~(T_{ab}-\frac{1}{2}Tg_{ab})-R_{ab}$. Then
by the Bianchi identities one gets
\begin{equation}g^{cb}\partial_{c}\Delta_{ab}=\nabla_{b}T^{b}_{a}-\frac{1}{2}\partial_{a}(T+R)+2g^{cb}\Gamma^{d}_{c(a}\Delta_{b)d}\end{equation}
\\
and note that~$T+R=-g^{ab}\Delta_{ab}$.
\\From (82)-(84)~$\Delta_{ab}$~is given by
\begin{equation}\Delta_{ij}=\partial_{(i}(\hat{\Gamma}_{j)}-\Gamma_{j)})+~\textrm{terms~in}~(\hat{\Gamma}_{a}-\Gamma_{a})\end{equation}
\begin{displaymath}\Delta_{iz}=\frac{1}{2}\left(\frac{1+2\gamma}{3\gamma-1}\right)\partial_{i}(\hat{\Gamma}_{z}-\Gamma_{z})+\left(\frac{3\gamma-2}{4\gamma}\right)\partial_{z}(\hat{\Gamma}_{i}-\Gamma_{i})\end{displaymath}
\begin{equation}+\textrm{terms~in}~(\hat{\Gamma}_{a}-\Gamma_{a})\end{equation}
\begin{displaymath}\Delta_{zz}=\frac{(5\gamma-1)}{2(3\gamma-1)}\partial_{z}(\hat{\Gamma}_{z}-\Gamma_{z})+\left(\frac{1-\gamma}{4V^{2}\gamma}\right)h^{ij}\partial_{i}(\hat{\Gamma}_{j}-\Gamma_{j})\end{displaymath}
\begin{equation}+~\mathrm{terms~in}~(\hat{\Gamma}_{a}-\Gamma_{a})\end{equation}
\\
Now substituting (85),(86), (88)-(90) into (87) gives
\begin{displaymath}g^{cb}\partial_{c}\Delta_{ib}=\frac{V^{2}}{2}\left(\frac{5\gamma-1}{2(3\gamma-1)}\right)\partial^{2}_{iz}(\hat{\Gamma}_{z}-\Gamma_{z})+\frac{1}{2}\left(\frac{1+3\gamma}{4\gamma}\right)h^{mn}\partial^{2}_{im}(\hat{\Gamma}_{n}-\Gamma_{n})
\end{displaymath}
\begin{displaymath}-\frac{3V^{2}(2-\gamma)(\gamma-1)}{(3\gamma-2)^{2}}\frac{(\hat{\Gamma}_{i}-\Gamma_{i})}{Z^{2}}+~\mathrm{terms~in}~(\hat{\Gamma}_{a}-\Gamma_{a}),~Z^{-1}(\hat{\Gamma}_{a}-\Gamma_{a}),~\partial_{b}(\hat{\Gamma}_{a}-\Gamma_{a})\end{displaymath}

\begin{equation}+\frac{V^{2}}{Z}\left(\frac{2}{3\gamma-2}\right)\left\{\left(\frac{3-4\gamma}{1-3\gamma}\right)\partial_{i}(\hat{\Gamma}_{z}-\Gamma_{z})+\left(\frac{3-4\gamma}{2\gamma}\right)\partial_{z}(\hat{\Gamma}_{i}-\Gamma_{i})\right\}\end{equation}
\\
\begin{displaymath}g^{cb}\partial_{c}\Delta_{zb}=\frac{V^{2}}{4}\left(\frac{5\gamma-1}{3\gamma-1}\right)\partial^{2}_{z}(\hat{\Gamma}_{z}-\Gamma_{z})+\left(\frac{1+3\gamma}{8\gamma}\right)h^{mn}\partial^{2}_{mz}(\hat{\Gamma}_{n}-\Gamma_{n})\end{displaymath}
\begin{displaymath}+\frac{V^{2}}{Z^{2}}(\hat{\Gamma}_{z}-\Gamma_{z})\left(\frac{2(13\gamma-8)-6(V^{P}-1)(5\gamma^{2}-4\gamma+3)}{(1-3\gamma)(3\gamma-2)^{2}}\right)\end{displaymath}
\begin{equation}+\mathrm{terms~in}~(\hat{\Gamma}_{a}-\Gamma_{a}),~Z^{-1}(\hat{\Gamma}_{a}-\Gamma_{a}),~\partial_{b}(\hat{\Gamma}_{a}-\Gamma_{a})\end{equation}
\\
The unspecified terms in these equations are polynomial in the stated
quantities, with coefficients which are polynomial in the components
of~$u$.\\\\
One can also calculate~$g^{cb}\partial_{c}\Delta_{ab}$~directly from
(88)-(90), giving
\begin{displaymath}g^{cb}\partial_{c}\Delta_{ib}=\frac{V^{2}}{2}\left(\frac{1+2\gamma}{3\gamma-1}\right)\partial^{2}_{iz}(\hat{\Gamma}_{z}-\Gamma_{z})+V^{2}\left(\frac{3\gamma-2}{4\gamma}\right)\partial^{2}_{z}(\hat{\Gamma}_{i}-\Gamma_{i})\end{displaymath}

\begin{displaymath}+\frac{1}{2}h^{mn}\partial^{2}_{mn}(\hat{\Gamma}_{i}-\Gamma_{i})+\frac{1}{2}h^{mn}\partial^{2}_{mi}(\hat{\Gamma}_{n}-\Gamma_{n})\end{displaymath}
\begin{equation}+\mathrm{terms~as~in~(91)}\end{equation}
\\
\begin{displaymath}g^{cb}\partial_{c}\Delta_{zb}=\frac{V^{2}}{4}\left(\frac{5\gamma-1}{3\gamma-1}\right)\partial^{2}_{z}(\hat{\Gamma}_{z}-\Gamma_{z})+\left(\frac{2\gamma-1}{4\gamma}\right)h^{mn}\partial^{2}_{mz}(\hat{\Gamma}_{n}-\Gamma_{n})\end{displaymath}
\begin{displaymath}+\frac{1}{2}\left(\frac{2\gamma+1}{3\gamma-1}\right)h^{mn}\partial^{2}_{mn}(\hat{\Gamma}_{z}-\Gamma_{z})\end{displaymath}
\begin{equation}+\mathrm{terms~as~in~(92)}\end{equation}
\\
Now we can compare (91) with (93) and (92) with (94) to obtain
\begin{displaymath}V^{2}\left(\frac{3\gamma-2}{4\gamma}
\right)\partial^{2}_{z}(\hat{\Gamma}_{i}-\Gamma_{i})=
\frac{V^{2}(\gamma-3)}{4(3\gamma-1)}\partial^{2}_{iz}
(\hat{\Gamma}_{z}-\Gamma_{z})+\left(\frac{1-\gamma}{8\gamma}\right)
h^{mn}\partial^{2}_{mi}(\hat{\Gamma}_{n}-\Gamma_{n})\end{displaymath}
\begin{displaymath}-\frac{1}{2}h^{mn}\partial^{2}_{mn}(\hat{\Gamma}_{i}-\Gamma_{i})+~\frac{V^{2}}{Z}\Bigg\{-\frac{3(2-\gamma)(\gamma-1)}{(3\gamma-2)^{2}}\left(\frac{\hat{\Gamma}_{i}-\Gamma_{i}}{Z}\right)\end{displaymath}
\begin{displaymath}+\frac{2}{(3\gamma-2)}\left(\left(\frac{3-4\gamma}{1-3\gamma}\right)\partial_{i}(\hat{\Gamma}_{z}-\Gamma_{z})+\left(\frac{3-4\gamma}{2\gamma}\right)\partial_{z}(\hat{\Gamma}_{i}-\Gamma_{i})\right)\Bigg\}
\end{displaymath}
\begin{equation}+\mathrm{terms~as~in~(91)}\end{equation}\\
\\

\begin{displaymath}\frac{V^{2}}{4}\left(\frac{5\gamma-1}{3\gamma-1}\right)\partial^{2}_{z}(\hat{\Gamma}_{z}-\Gamma_{z})=\frac{(3-\gamma)}{8\gamma}h^{mn}\partial^{2}_{mz}(\hat{\Gamma}_{n}-\Gamma_{n})\end{displaymath}
\begin{displaymath}-\frac{1}{2}\left(\frac{1+2\gamma}{3\gamma-1}\right)h^{mn}\partial^{2}_{mn}(\hat{\Gamma}_{z}-\Gamma_{z})+~\mathrm{terms~as~in~(93)}\end{displaymath}
\begin{equation} +\frac{V^{2}}{Z}\left(\frac{\hat{\Gamma}_{z}-\Gamma_{z}}{Z}\right)\left(\frac{2(13\gamma-8)-6(V^{P}-1)(5\gamma^{2}-4\gamma+3)}{(1-3\gamma)(3\gamma-2)^{2}}\right)\end{equation}
\\
Next we introduce new variables as follows
\begin{displaymath}\pi_{i}=\hat{\Gamma}_{i}-\Gamma_{i},~~\tau=\hat{\Gamma}_{z}-\Gamma_{z},~~\alpha_{ij}=\partial_{i}(\hat{\Gamma}_{j}-\Gamma_{j}),~~\beta_{i}=\partial_{z}(\hat{\Gamma}_{i}-\Gamma_{i})\end{displaymath}
\begin{displaymath}\omega_{i}=C_{1}\partial_{i}(\hat{\Gamma}_{z}-\Gamma_{z}),~~\delta=C_{2}\partial_{z}(\hat{\Gamma}_{z}-\Gamma_{z}),~~\eta_{i}=Z^{-1}(\hat{\Gamma}_{i}-\Gamma_{i}),~~\kappa=Z^{-1}(\hat{\Gamma}_{z}-\Gamma_{z})\end{displaymath}
where
\begin{displaymath}C_{1}=C_{2}\sqrt{\frac{2\gamma+1}{2(3\gamma-1)}},~~~C_{2}=\sqrt{\frac{2\gamma}{3\gamma-1}}\end{displaymath}
In terms of these new variables (95) becomes
\begin{displaymath}-V^{2}\left(\frac{3\gamma-2}{4\gamma}\right)h^{ij}\partial_{z}\beta_{i}=-\frac{(\gamma-3)}{4(3\gamma-1)}\frac{V^{2}}{C_{2}}h^{ij}\partial_{i}\delta-\left(\frac{1-\gamma}{8\gamma}\right)h^{ij}h^{kn}\partial_{k}\alpha_{in}\end{displaymath}
\begin{displaymath}+\frac{1}{2}h^{ij}h^{km}\partial_{k}\alpha_{mi}+\mathrm{~order~zero~terms}\end{displaymath}
\begin{equation}-\frac{V^{2}}{Z}h^{ij}\left\{-\frac{3(2-\gamma)(\gamma-1)}{(3\gamma-2)^{2}}\eta_{i}+\frac{2(3-4\gamma)}{(3\gamma-2)(1-3\gamma)C_{1}}\omega_{i}+\frac{(3-4\gamma)}{\gamma(3\gamma-2)}\beta_{i}\right\}\end{equation}
and (96) becomes
\begin{displaymath}\frac{V^{4}}{4}\left(\frac{5\gamma-1}{3\gamma-1}\right)\partial_{z}\delta=V^{2}C_{2}\left(\frac{3-\gamma}{8\gamma}\right)h^{mn}\partial_{m}\beta_{n}-\frac{V^{2}C_{2}}{2C_{1}}\left(\frac{2\gamma+1}{3\gamma-1}\right)h^{mn}\partial_{m}
\omega_{n}\end{displaymath}
\begin{displaymath}+\frac{V^{4}}{Z}\left(\frac{2(13\gamma-8)-6(V^{P}-1)(5\gamma^{2}-4\gamma+3)}{(1-3\gamma)(3\gamma-2)^{2}}\right)\kappa\end{displaymath}
\begin{equation}+~\mathrm{order~zero~terms}\end{equation}
\\
By the above definitions one also obtains the following
\begin{equation}\partial_{z}\pi_{i}=~\beta_{i}\end{equation}
\begin{equation}\partial_{z}\tau=~(C_{2})^{-1}\delta\end{equation}
\begin{displaymath}\left(\frac{1}{2}h^{im}h^{jn}+\left(\frac{\gamma-1}{8\gamma}\right)h^{in}h^{jm}\right)\partial_{z}\alpha_{mn}=~\left(\frac{1}{2}h^{im}h^{jn}+\left(\frac{\gamma-1}{8\gamma}\right)h^{in}h^{jm}\right)\partial_{m}\beta_{n}
\end{displaymath}
\begin{equation} \end{equation}
\begin{equation}-V^{2}h^{ij}\partial_{z}\omega_{j}=~-\frac{V^{2}C_{1}}{C_{2}}h^{ij}\partial_{j}\delta\end{equation}
\begin{equation}-V^{2}h^{ij}\partial_{z}\eta_{j}=~-\frac{V^{2}}{Z}h^{ij}(-\eta_{j}+\beta_{j})\end{equation}
\begin{equation}V^{4}\partial_{z}\kappa=~\frac{V^{4}}{Z}((C_{2})^{-1}\delta-\kappa)\end{equation}

The system (97)-(104) is of the form (see Appendix):
\begin{equation}a^{0}(v)\partial_{z}v=~a^{i}(v)\partial_{i}v+~b(v)v+~\frac{1}{z}c(v)v\end{equation}
where ~$v$~stands for the variables written as a vector, ~$a^{0}$~is
symmetric and
positive definite, the~$a^{i}$~are 
symmetric, and all coefficients are polynomial in the components
of~$v$~and those of~$u$~(~$u$~as in (70))
\\
The eigenvalues of~$(a^{0})^{-1}c(0)$~satisfy one of
\begin{displaymath}\lambda=0~,~~\lambda^{2}+\lambda+\frac{8(13\gamma-8)}{(5\gamma-1)(3\gamma-2)^{2}}=0\end{displaymath}
\begin{displaymath}(1+\lambda)\left(\lambda+\frac{4(4\gamma-3)}{(3\gamma-2)^{2}}\right)+~\frac{12\gamma(2-\gamma)(\gamma-1)}{(3\gamma-2)^{3}}=0\end{displaymath}
Hence for ~$1<\gamma<2$, no~$\lambda$~is a positive integer.
\\\\
The initial data for (97)-(104) is as follows: One
has~$\hat{\Gamma}_{i}=0$~at~$Z=0$. Since~$\phi$~is constant
at~$Z=0$~one also has ~$\Gamma_{i}=0$~there. Hence~$\pi_{i}=~\alpha_{ij}=0$~
initially.
Now~$\hat{\Gamma}_{z}=\nu=0$~at~$Z=0$.
Since~$k_{ij}=0$~at~$Z=0$,~(58)
implies~$\partial_{z}V=0$~there. Hence~$\Gamma_{z}=0$~at~$Z=0$~and
so~$\tau=~\omega_{i}=0$~there.\\
Taking a Taylor expansion of (61), (64) gives
~$\partial_{z}\xi_{i}=~\partial_{z}\hat{\Gamma}_{i}=0$~at~$Z=0$. The
identity (51) then gives~$\partial_{z}\Gamma_{i}=0$~at~$Z=0$, and so
~$\beta_{i}=0$~there.
\\
A Taylor expansion of (59), (63), (58) gives
~$\partial^{2}_{z}V=\frac{1}{P}\partial_{z}\hat{\Gamma}_{z}$~at~$Z=0$~.
Comparison with (72) then
gives~$\partial_{z}(\hat{\Gamma}_{z}-\Gamma_{z})=0$~at~$Z=0$. Thus~$\delta=~\kappa=0$~at~$Z=0$.\\

All the initial data for (97)-(104) are therefore zero, and one
solution of this system is~$v\equiv 0$. For sufficiently smooth~$v$~this
solution is unique by Theorem 2, and we have
recovered~$\hat{\Gamma}_{a}=\Gamma_{a}$~(An analogue of Theorem 2
can be obtained from Theorem 1 in the case when certain of the
coefficients are polynomial in suitably regular known functions). The metric~$g$~obtained
from~$u$~in (70) via (38) is thus a
solution of the conformal Einstein-perfect fluid equations (22).\\

We now discuss the differentiability of solutions to (56)-(68)
and (97)-(104). Suppose then that~$\Sigma$~is a smooth paracompact 
manifold. In order
to apply Theorem 1 one must first consider the localised problem, and then
use the finite speed of propagation inherent in (70), (105) to
globalise (see Claudel and Newman 1998). Specifically, choose a 
locally finite open
cover~$\{O_{\alpha}\}$~of harmonic coordinate charts for the
metric~$h_{ij}$~on~$\Sigma$, with each~$O_{\alpha}$~having compact
closure therein. And let~$\{O'_{\alpha}\}$~be another cover
for~$\Sigma$~satisfying~$\bar{O}'_{\alpha}\subset O_{\alpha}$. Suppose
that the initial metric lies in the following Sobolev class
\begin{equation}(h_{ij})_{\alpha}\in~H^{2q_{\alpha}+2+s}~~,~~s>3/2\end{equation}
where
\begin{equation}q_{\alpha}>\mathrm{max}\{3+\|(a^{0})^{-1}c\|_{B(\mathbb{R}^{24})},~1+\|(A^{0})^{-1}F(0,x)\|_{B(\mathbb{R}^{63})}\}~,~x\in~O_{\alpha}\end{equation}
Then Theorem 2 gives that there is a solution~$(u)'_{\alpha}$~of
(56)-(58) contained and unique in the following class
\begin{equation}(u)'_{\alpha}\in~\bigcap_{i=0}^{q_{\alpha}}C^{i}([0,T_{q_{\alpha}}],~H^{q_{\alpha}+s-i}(O'_{\alpha}))\end{equation}
\\
Now by the definition of~$v$~in (105) there follows
\begin{equation}(v)'_{\alpha}\in~\bigcap_{i=0}^{(q_{\alpha}-2)}C^{i}([0,T_{q_{\alpha}}],H^{q_{\alpha}+s-2-i}(O'_{\alpha}))\end{equation}
and (an analogue of) Theorem 2 gives that~$v\equiv 0$~. Hence the metric~$(g_{ab})'_{\alpha}$~obtained from
~$(u'_{\alpha})$~is a solution of the conformal Einstein-perfect fluid
equations (22), and one has
\begin{equation}(g_{ab})'_{\alpha}\in~\bigcap_{i=0}^{q_{\alpha}}C^{i}([0,T_{q_{\alpha}}],H^{q_{\alpha}+s-i}(O'_{\alpha}))\end{equation}
with~$(g_{ab})'_{\alpha}$~unique in this class.\\
\\Note that if we give ~$C^{\infty}$~ data~$(h_{ij})_{\alpha}$~at~$\Sigma$~then the
evolution given by (108) is ~$C^{\infty}$~on some time
interval~$[0,T_{\alpha}]$. For the data appropriate to the localised problem
is~$C_{0}^{\infty}$~and thus one gets~$(u)'_{\alpha}$~as in (108) for arbitrarily
large~$q_{\alpha}$. For fixed ~$\alpha$~the ~$T_{q_{\alpha}}$~are positively bounded below by the
following standard extension argument: first from (108) one gets by the
Sobolev imbedding theorem (Adams 1975) that
~$(u)'_{\alpha}$~is~$C^{1}$~on~$[0,T_{\alpha}]$~for some~$T_{\alpha}>0$. Suppose now that~$(u)'_{\alpha}\in
C([0,T_{1}),H^{r})$~for some~$r>\frac{5}{2},~T_{1}<T_{\alpha}$. Then, by
proposition 5.1.E of (Taylor 1991),~$\|(u)'_{\alpha}\|_{H^{r}}$~is bounded on~$[0,T_{1})$~and~$(u)'_{\alpha}$~can
be uniquely extended in~$H^{r}$~onto~$[0,T_{1}]$. Then standard
theory for regular symmetric hyperbolic PDE gives an extension
of~$(u)'_{\alpha}$~onto a larger interval~$[0,T_{2})$~with
\begin{displaymath}(u)'_{\alpha}\in\bigcap_{i=0}^{r}C^{i}([T_{1},T_{2}),H^{s-i}(O'_{\alpha}))\end{displaymath}
Thus~$(u)'_{\alpha}$~can be extended onto~$[0,T_{\alpha}]$~in the above way. It follows
by the Sobolev imbedding theorem that~$(u)'_{\alpha}$~is~$C^{\infty}$~on~$[0,T_{\alpha}]$.

\section{Solving the Cauchy problem for~$\gamma=2$}
The method of section (4) is clearly inadequate for dealing with the
cases\linebreak $\gamma=1$ and $\gamma= 2$ as certain of the coefficients in the field
equations diverge at these values of the polytropic index. However, in
the case ~$\gamma=2$~one must have~$\square Z=0$~by (26), (32),
and it turns out that the problem may be successfully attacked using
harmonic coordinates in~$M$. Unfortunately in the case~$\gamma=1$~it
has not been possible to write the field equations in symmetric
hyperbolic form, and the problem has not been solved in such
generality. However in section 6 we are able to solve the problem in 
the case of Bianchi type spatial homogeneity.
\\\\
For the case~$\gamma=2$~first note that the conformal field equations are just
\begin{equation}\nabla_{a}\nabla_{b}Z=ZR_{ab}\end{equation}
Again we wish to solve the Cauchy problem
on~$M=\Sigma\mathbf{\times}[0,T]$. We will use
coordinates~$x^{0}=Z,~x^{i}$,~i=1,2,3~on~$M$, and in the sequel Greek
indices will take the values 0, 1, 2, 3.
\\\\
Now let~$H^{\alpha}=\square x^{\alpha}$, and consider the following
reduced equations for ~$g_{\mu\nu}$
\begin{equation}ZR^{H}_{\mu\nu}=\nabla_{\mu}\nabla_{\nu}Z-\frac{H^{0}}{g^{00}}\Delta_{\mu\nu}\end{equation}
where
\begin{equation}R^{H}_{\mu\nu}=R_{\mu\nu}+~g_{\alpha(\mu}\partial_{\nu)}H^{\alpha}\end{equation}
\begin{equation}\Delta_{ij}=\Delta_{i0}=\Delta_{0i}=0,~~~\Delta_{00}=1\end{equation}
Note that (112) implies$~R^{H}=0$~so
that~$G^{H}_{\mu\nu}=R^{H}_{\mu\nu}$
\\Now
write~$h_{\alpha\beta\gamma}=\partial_{\gamma}g_{\alpha\beta}$. Then
(112) can be expressed as follows
\begin{equation}-g^{(ij)}\partial_{z}h_{\alpha\beta
j}=-g^{(ij)}\partial_{j}h_{(\alpha\beta)0}\end{equation}
\begin{equation}\partial_{z}g_{\alpha\beta}=h_{\alpha\beta
0}\end{equation}
\begin{equation}\partial_{z}g^{\alpha\beta}=-g^{\alpha\mu}g^{\beta\nu}h_{\mu\nu
0}\end{equation}
\begin{displaymath}g^{00}\partial_{z}h_{\alpha\beta
0}=-g^{(ij)}\partial_{i}h_{(\alpha\beta)j}-2g^{0i}\partial_{i}h_{(\alpha\beta)0}\end{displaymath}
\begin{equation}+2H_{(\alpha\beta)}+\frac{1}{Z}g^{0\sigma}(-h_{\alpha\beta\sigma}+h_{\beta\sigma\alpha}+h_{\alpha\sigma\beta})\end{equation}
(for~$\alpha,~\beta$~not both zero)
\begin{displaymath}g^{00}\partial_{z}h_{000}=-g^{(ij)}\partial_{i}h_{00j}-2g^{0i}\partial_{i}h_{000}\end{displaymath}
\begin{equation}+2H_{00}+\frac{1}{Z}\{\frac{1}{g^{00}}g^{ij}g^{0k}(h_{ijk}-2h_{ikj})+\frac{2}{g^{00}}g^{0k}g^{0i}(h_{0ki}-h_{0ik}-h_{ki0})+Q\}\end{equation}
where
\begin{equation}Q=g^{ij}(h_{ij0}-2h_{i0j})-2g^{0i}h_{00i}\end{equation}
We can write the singular part of (118) as
~$\frac{1}{Z}C_{1}(u)u$~in such a way that the components of~$C_{1}(u)u$~are
\begin{equation}\tilde{g}^{00}(h_{\beta 0\alpha}+h_{\alpha
0\beta}-h_{\alpha\beta 0})+(\tilde{h}_{\beta
i\alpha}+\tilde{h}_{\alpha i\beta}-\tilde{h}_{\alpha\beta
i})g^{0i}\end{equation}
where the tilded terms are in ~$C_{1}(u)$~and the untilded terms are
in~$u$. And the singular part of (119) can be written
as~$\frac{1}{Z}C_{2}(u)u$~with components
\begin{displaymath}\frac{1}{\tilde{g}^{00}}(\tilde{h}_{ijk}-2\tilde{h}_{ikj})\tilde{g}^{ij}g^{0k}+\frac{2}{\tilde{g}^{00}}(\tilde{h}_{0ki}-\tilde{h}_{0ik}-\tilde{h}_{ki0})\tilde{g}^{0i}g^{0k}\end{displaymath}
\begin{equation}+\tilde{g}^{ij}(h_{ij0}-2h_{i0j})-2\tilde{g}^{0i}h_{00i}\end{equation}
where the tilded terms are in~$C_{2}(u)$~and the untilded terms are
in~$u$.
\\Clearly equations (115)-(119) are of the form:
\begin{equation}A^{0}(u)\partial_{z}u=~A^{i}(u)\partial_{i}(u)+B(u)u+\frac{1}{Z}C(u)u\end{equation}
where~$u$~stands for the field variables written as a vector,
the~$A^{\alpha}$~are symmetric with~$A^{0}$~positive definite, and the
components of~$C$~are those of~$C_{1},~C_{2}$~written in an
appropriate order.
\subsection{Initial data for the reduced equations}
For the initial data we first choose a negative definite
3-metric~$h_{ab}$~on~$\Sigma$. Next choose coordinates
~$x^{i}$~on~$\Sigma$~satisfying the harmonic gauge
condition~$h^{jk}\Gamma^{i}_{jk}=0$, and also take $~g_{0i}=0$. We know
from the field equations that ~$g^{00}=1,~h_{ij0}=0$~initially, and
hence~$g_{00}=1$~initially. We make the choices~$h_{\alpha 00}(0)=0$, and this
ensures~$\square x^{\alpha}=0$~at~$\Sigma$. These choices also ensure
that ~$(C(u))u(0)=0~\forall u$.
\subsection{The eigenvalues of~$(A^{0})^{-1}C(0)$}
The eigenvalue equation for~$(A^{0})^{-1}C(0)$~amounts to the following
\begin{displaymath}
\lambda h_{\alpha\beta 0}=(h_{\beta 0\alpha}+h_{\alpha
0\beta}-h_{\alpha\beta 0})+\end{displaymath}\begin{displaymath}g^{0i}(\bar{h}_{\beta
i\alpha}+\bar{h}_{\alpha i\beta}-\bar{h}_{\alpha\beta i})\end{displaymath}
(with~$h_{000}$~replaced by~$\bar{g}^{ij}(h_{ij0}-2h_{i0j})$~in rhs)
\begin{displaymath}\lambda h_{\alpha\beta
k}=0\end{displaymath}
\begin{displaymath}\lambda g^{0i}=0\end{displaymath}
(Unbarred quantities are elements of an eigenvector, barred quantities
are evaluated at~$\Sigma$~)
\\So suppose~$\lambda\neq0$. Then 1 implies
\begin{displaymath}\lambda
h_{ij0}=-h_{ij0}+g^{0m}\{\bar{h}_{jmi}+\bar{h}_{imj}-\bar{h}_{ijm}\}\end
{displaymath}
Since the ~$x^{i}$~are harmonic in~$\Sigma$~this implies
\begin{displaymath}\lambda
\bar{g}^{ij}h_{ij0}=-\bar{g}^{ij}h_{ij0}\end{displaymath}
\\From 1 one also gets
\begin{displaymath}\lambda h_{000}=\bar{g}^{ij}h_{ij0}~~,~~\lambda
h_{i00}=0\end{displaymath}
It follows that~$\lambda=0$~or~$-1$. The system
(115)-(119) therefore satisfies all the requirements of Theorem 2, and has a solution ~$u$~for each choice of
~$h_{ab}$~on~$\Sigma$, unique in a suitable differentiability
class. Our next aim is to show, as in section (19), that the
4-metric~$g_{ab}$~obtained from~$u$~is actually a solution of the full field
equations (111).
\subsection{Recovering the Einstein equations}
Firstly we show that if (115)-(119) are satisfied, then
~$g_{\alpha\beta}$~is symmetric and the definition~$h_{\alpha\beta
j}=\partial_{j}g_{\alpha\beta}$~can be recovered.\\
Observe that (118)-(119) imply
\begin{displaymath}g^{00}\partial_{z}h_{[\alpha\beta]0}=-\frac{1}{Z}g^{00}h_{[\alpha\beta]0}-\frac{1}{Z}g^{0i}h_{[\alpha\beta]i}\end{displaymath}
and (115) implies
\begin{displaymath}\partial_{z}h_{[\alpha\beta]j}=0\end{displaymath}
Thus~$h_{[\alpha\beta]j}=0$~, and there follows
\begin{displaymath}\partial_{z}(Zh_{[\alpha\beta]0})=0\end{displaymath}
Hence~$h_{[\alpha\beta]0}\equiv 0$~, and now (2.101)
implies~$g_{\alpha\beta}=g_{\beta\alpha}$.
\\From (115) one now gets
\begin{displaymath}\partial_{z}(h_{\alpha\beta
j}-\partial_{j}g_{\alpha\beta})=
\partial_{j}h_{\alpha\beta
0}-\partial^{2}_{zj}g_{\alpha\beta}=0\end{displaymath}
by (116), so that ~$h_{\alpha\beta j}=\partial_{j}g_{\alpha\beta}$.

We will now use the Bianchi identities to show that if equations
(115)-(119) (or equivalently (112)) are satisfied then we have
~$\square x^{\alpha}=0$~, and thus (111) is solved.
 
So suppose (115)-(119) hold. Then one calculates
\begin{displaymath}\nabla^{\mu}R^{H}_{\mu\nu}=-\frac{1}{Z}\Big[(\nabla^{\mu}Z)(R^{H}_{\mu\nu}-R_{\mu\nu})-\partial_{\nu}H^{0}\end{displaymath}
\begin{equation} +g^{\mu\rho}\left\{\Delta_{\mu\nu}\partial_{\rho}(H^{0}/g^{00})-(H^{0}/g^{00})(\Gamma^{0}_{\rho\mu}\Delta_{\nu
0}+\Gamma^{0}_{\rho\nu}\Delta_{0\mu})\right\}\Big]\end{equation}
The Bianchi identities now give
\begin{displaymath}0=g^{00}\partial^{2}_{z}H^{\alpha}+2g^{0i}\partial^{2}_{0i}H^{\alpha}+g^{ij}\partial^{2}_{ij}H^{\alpha}\end{displaymath}
\begin{displaymath}+\frac{1}{Z}\Big\{g^{\gamma
0}\partial_{\gamma}H^{\alpha}-g^{\nu\alpha}\partial_{\nu}H^{0}+2g^{0\alpha}g^{0\rho}((g^{00})^{-1}\partial_{\rho}H^{0}-(g^{00})^{-2}H^{0}h_{00\rho})\end{displaymath}
\begin{equation}+2(g^{0\alpha}/g^{00})(H^{0})^{2}-2g^{\nu\alpha}g^{0\rho}(H^{0}/g^{00})\Gamma^{0}_{\rho\nu}\Big\}\end{equation}

If we now put~$h_{\alpha}^{\beta}=\partial_{\alpha}H^{\beta}$~, then
(125) can be written in first order form as follows
\begin{equation}-g^{ij}\partial_{z}h_{j}^{\alpha}=-g^{ij}\partial_{j}h_{0}^{\alpha}\end{equation}
\begin{equation}\partial_{z}H^{\alpha}=h_{0}^{\alpha}\end{equation}
\begin{displaymath}g^{00}\partial_{z}h_{0}^{\alpha}=-2g^{0i}\partial_{i}h_{0}^{\alpha}-g^{ij}\partial_{i}h_{j}^{\alpha}\end{displaymath}
\begin{displaymath}+\frac{1}{Z}\Big\{-g^{0\gamma}h_{\gamma}^{\alpha}+g^{\nu\alpha}h_{\nu}^{0}+2g^{0\alpha}g^{0\rho}(-(h_{\rho}^{0}/g^{00})+(g^{00})^{-2}H^{0}h_{00\rho})\end{displaymath}
\begin{equation}-2(g^{0\alpha}/g^{00})(H^{0})^{2}+2g^{\nu\alpha}g^{0\rho}(H^{0}/g^{00})\Gamma^{0}_{\rho\nu}\Big\}\end{equation}

This system is symmetric hyperbolic:
\begin{equation}a^{0}(v)\partial_{z}v=a^{i}(v)\partial_{i}v+b(v)v+\frac{1}{Z}c(v)v\end{equation}
where ~$v$~stands for~$H^{\alpha},~h_{\alpha}^{\beta}$. We know
that~$H^{\alpha}=0$~at~$\Sigma$~, and now equation (128) implies
that~$h_{0}^{\alpha}=0$~at~$\Sigma$~also. Hence all the initial data
for (126)-(128) are zero, and one solution is trivial.
\\\\
The eigenvalues of ~$c(0)$~satisfy either~$\lambda=0$~or
\begin{displaymath}\lambda_{0}^{\alpha}=-h_{0}^{\alpha}+\bar{g}^{\nu\alpha}h^{0}_{\nu}-2\bar{g}^{0\alpha}\bar{g}^{0\rho}h^{0}_{\rho}\end{displaymath}
Putting~$\alpha=0, j$~gives, respectively
\begin{displaymath}\lambda h_{0}^{0}=-2h_{0}^{0}~~~,~~~\lambda
h_{0}^{i}=-h_{0}^{i}\end{displaymath}
Hence~$\lambda\in\{0, -1, -2\}$. So  (126)-(128) is in Newman-Claudel
form, and the zero solution is unique in a suitable differentiability
class, so that (111) is satisfied.
\subsection{Existence, uniqueness, and differentiability}
The situation here is very similar to that in section (4). So take
covers~$\{O_{\alpha}\},~\{O'_{\alpha}\}$~for~$\Sigma$~as before, and
suppose that the initial 3-metric lies in the following class
\begin{equation}(h_{ij})_{\alpha}\in~H^{2q_{\alpha}+2+s}(O_{\alpha})~~,~~s>3/2\end{equation}
where
\begin{equation}q_{\alpha}>~\mathrm{max}\{3+\|(a^{0})^{-1}c(0,x)\|_{B(\mathbb{R}^{96})}~,~1+\|(A^{0})^{-1}C(0,y)\|_{B(\mathbb{R}^{20})}\}~~~x,
y\in O_{\alpha}\end{equation}
Then by Theorem 2 there exists a solution~$u$~of (112) contained and unique in
the following class
\begin{equation}(u)'_{\alpha}\in~\bigcap_{i=0}^{q_{\alpha}}C^{i}([0,T_{q_{\alpha}}],
H^{q_{\alpha}+s-i}(O'_{\alpha}))\end{equation}
\\From the definition of~$H^{\alpha},~h_{\alpha}^{\beta}$~it follows
that
\begin{equation}(v)'_{\alpha}\in~\bigcap_{i=0}^{(q_{\alpha}-2)}C^{i}([0,T_{q_{\alpha}}],~H^{q_{\alpha}+s-2-i}(O'_{\alpha}))\end{equation}
and by (an analogue of) Theorem 2 one must have~$v\equiv 0$. Hence the metric
~$g_{ab}$~obtained from~$u$~is the unique solution of the Einstein
equations (111) with starting metric~$h_{ij}$~on~$\Sigma$.
\\
We note that, as in section 4,~$C^{\infty}$~data gives rise to a~$C^{\infty}$~solution~$u$.
\subsection{The Vanishing Weyl Tensor Conjecture}
The results of sections 4-5 can now be used to prove the Vanishing
Weyl Tensor conjecture. From (Goode and Wainwright 1985), and as one 
can see from equation (9) in this case, one knows that the Weyl 
curvature vanishes at
~$\Sigma$~iff the initial 3-metric is of constant curvature, and that
any such metric is initial data for some FRW cosmology. The 4-metric
of any FRW cosmology is a smooth evolution of smooth data, and hence
by the above is the only smooth evolution of the same data. Thus in
this case the conjecture is true: for~$1<\gamma\leq 2$~at least the
only cosmologies with an isotropic singularity and vanishing initial
Weyl tensor are the FRW
models where the Weyl tensor vanishes everywhere.\\

\section{Spatially homogeneous~$\gamma=1$~spacetimes}
We noted above that for~$\gamma=1$~the conformal
Cauchy problem has not been solved in full generality. However if one restricts attention to
certain (Bianchi type) spatially homogeneous cosmologies, the field
equations become ODE's and a theorem of Rendall and Schmidt (1991) can
be used to obtain the required existence and uniqueness result, as we
shall see in this section.

\subsection{Bianchi type spacetimes}
The Bianchi type spacetimes are a class of spatially homogeneous
cosmological models having a 3-parameter Lie group G of
isometries. They are classified into 9 types according to the nature
of the Lie algebra associated with G (see Wald 1984 for details). In such spacetimes there
exists a cosmic time function~$t$, and covector
fields~$(e^{i})_{a}$~such that the metric can be written as
\begin{equation}\tilde{g}_{ab}=\nabla_{a}t\nabla_{b}t+\tilde{h}_{ij}(t)(e^{i})_{a}(e^{j})_{b}\end{equation}
The ~$(e^{i})_{a}$~are preserved under the action of G, and are
called~\textit{left-invariant}~one-form fields.
\subsection{Conformal field equations in~$M$}
We now consider equations (22), (37) in the case where the physical
metric ~$\tilde{g}_{ab}$~has Bianchi type symmetry. We may assume wlog that the singularity surface is at Bianchi time~$t=0$. One must also have that the fluid velocity is orthogonal to the homogeneous hypersurfaces, by the following argument: There exist
 vector fields~$(\xi_{i})^{a}~(i=1,2,3)$~on the Lie group~$G$~which are Killing vector fields for~$\tilde{g}_{ab}(t),~t>0$. Since the fluid velocity~$\tilde{u}^{a}$~is geodesic for~$\gamma=1$~there follows~$\tilde{u}_{a}(\xi_{i})^{a}=$~const. in~$\tilde{M
}$. But~$\tilde{u}_{a}=\Omega u_{a}$~, with~$u_{a}$~regular. Letting~$\Omega\rightarrow 0$~we see that~$\tilde{u}_{a}(\xi_{i})^{a}=0$~in~$\tilde{M}$~as required.\\

The conformal factor is
~$\Omega=Z^{2}$~and the unphysical metric can be written
\begin{equation}g_{ab}=\nabla_{a}Z\nabla_{b}Z+h_{ij}(Z)(e^{i})_{a}(e^{j})_{b}\end{equation}
since~$V=1$. It follows that~$\Omega=(3t)^{2/3}$~where~$t$~is the 
Bianchi time.\\

Suppose now that we write equations (22), in an obvious way, as~$G_{ab}=T_{ab}$. In order
that these equations be  satisfied at~$Z=0$~one must
have~$G(0)=1$~and~$K_{ij}(0)=0$~where~$K_{ij}\equiv\frac{1}{2}\partial_{z}h_{ij}$.
The initial 3-metric~$h_{ij}(0)$~is free data, as for the other
polytropes.\\

The equations~$G_{ij}=T_{ij}$~take the form
\begin{displaymath}\partial_{z}K_{ij}={}^{3}R_{ij}-KK_{ij}+2K_{im}K_{j}^{m}\end{displaymath}
\begin{equation}-\frac{4}{Z}K_{ij}-h_{ij}\Big(\frac{2K}{Z}+\frac{6(1-G)}{Z^{2}}\Big)\end{equation}
\begin{equation}\partial_{z}h_{ij}=2K_{ij}\end{equation}
where~$K=h^{ij}K_{ij}$~and~${}^{3}R_{ij}$~is the Ricci tensor of the
surfaces~$Z=$~constant, given by
\begin{displaymath}{}^{3}R_{ij}=\frac{1}{2}C^{m}_{mk}(C^{r}_{cj}h_{ir}+h_{jr}C^{r}_{ci})h^{kc}-\frac{1}{2}C^{c}_{ki}(C^{k}_{cj}+h_{cm}h^{kl}C^{m}_{lj})\end{displaymath}
\begin{equation}-\frac{1}{4}C^{m}_{sk}C^{r}_{lc}h_{jm}h_{ir}h^{kl}h^{cs}\end{equation}
where~$C^{i}_{jk}$~are the structure constants of the Lie group~$G$.
\\Now write
\begin{equation}C=(G_{ab}-T_{ab})Z^{a}Z^{b}\end{equation}
and
\begin{equation}C_{d}=(G_{ab}-T_{ab})h^{a}_{~d}Z^{b}\end{equation}
A rather lengthy calculation then shows that if one has a smooth
solution ~$h_{ij}(Z)$~of (136)-(137), then the following hold:
\begin{equation}\partial_{z}C=-D^{m}C_{m}-C\Big(K+\frac{4}{Z}\Big)\end{equation}
\begin{equation}\partial_{z}C_{i}=-C_{i}\Big(K+\frac{4}{Z}\Big)\end{equation}
where~$D_{i}$~is the derivative operator associated with~$h_{ij}$.
\\Now~$C$~and~$C_{i}$~vanish at~$Z=0$~by the conditions already
imposed. If we regard ~$K(Z)$~as a known smooth function then Theorem
1 of (Rendall and Schmidt 1991)
implies that (142) has a unique solution~$C_{i}$. Hence~$C_{i}$~is identically zero, and then in the same
way~$C$~is identically zero. Our task is therefore to solve equations
(37) and (136)-(137).\\

It will be convenient to work with ~$h_{ij},h^{ij},G,K_{i}^{~j}$~as
independent variables, and we also introduce the quantity~$\zeta$~via
\begin{equation}\zeta\equiv 6(1-G)Z^{-1}\end{equation}
The field equations to be solved are then
\begin{equation}\partial_{z}h_{ij}=2h_{(j|k|}K_{i)}^{~k}\end{equation}
\begin{equation}\partial_{z}h^{ij}=-2h^{(i|k|}K_{k}^{~j)}\end{equation}
\begin{equation}\partial_{z}K_{i}^{~j}=h^{jk}({}^{3}R_{ik})-K_{m}^{~m}K_{i}^{~j}-\frac{4}{Z}K_{i}^{~j}-\frac{1}{Z}\delta_{i}^{~j}(2K_{m}^{~m}+\zeta)\end{equation}
\begin{equation}\partial_{z}G=-GK_{m}^{~m}\end{equation}
\begin{equation}\partial_{z}\zeta=-\zeta K_{m}^{~m}+\frac{1}{Z}(6K_{m}^{~m}-\zeta)\end{equation}
Equations (144)-(148) can be written in matrix form as
\begin{equation}\frac{du}{dZ}+\frac{1}{Z}Nu=G(u)\end{equation}
where~$u=(h_{ij},h^{ij},G,K_{i}^{~j},\zeta)$,~$N$~is a constant
matrix, and~$G$~is polynomial in the components of~$u$.
\\The matrix~$N$~takes the form
\begin{equation}N=\left(\begin{array}{ll}O&O\\O&M\end{array}\right)\end{equation}
where the O's stand for blocks of zeros of various sizes and~$M$~is a
square matrix.\\
The eigenvalues ~$\lambda$~of~$M$~satisfy the following
\begin{equation}-\lambda
K_{i}^{~j}=-4K_{i}^{~j}-\delta_{i}^{~j}(2K_{m}^{~m}+\zeta)\end{equation}
\begin{equation}-\lambda\zeta=6K_{m}^{~m}-\zeta\end{equation}
Taking the trace of (151) gives
\begin{equation}(-\lambda+10)K_{m}^{~m}=-3\zeta\end{equation}
and  then from (152) there follows
\begin{equation}(-\lambda+10)(-\lambda+1)+18=0,~~\textrm{or}~~\zeta=0\end{equation}
Thus all the eigenvalues of~$M$~have strictly positive real part. Now if
all the eigenvalues of ~$N$~had strictly positive real parts, then 
Theorem 1 of (Rendall and Schmidt 1991) would imply the existence of a unique
smooth solution~$u$~to equations (149) with the given data. However,
in
the case where~$N$~takes the form (150), with~$M$~having eigenvalues
with strictly +ve real part,  a very elementary modification of their
proof gives the same result. It follows that equations (149) have a
unique smooth solution~$u$.                 
\\It remains to show
that if (149) is satisfied then~$h_{ij}=h_{ji},~h^{ij}h_{jk}=\delta^{i}_{~k}$,~$K_{ij}\equiv
h_{jk}K_{i}^{~k}=\frac{1}{2}\partial_{z}h_{ij}$, and~$\zeta=6Z^{-1}(1-G)$.\\
First note that (144), (145) imply
that~$\partial_{z}h_{[ij]}=\partial_{z}h^{[ij]}=0$, so
that~$h_{ij},~h^{ij}$\linebreak[4] stay symmetric if they start that way. Also
(144)-(145) imply
\begin{equation}\partial_{z}(h^{ij}h_{jk}-\delta^{i}_{~k})=(h^{ij}h_{jm}-\delta^{i}_{~m})K_{k}^{~m}-(h^{jm}h_{mi}-\delta^{m}_{~k})K_{m}^{~i}\end{equation}
Hence by Gronwall's
inequality~$h^{ij}h_{jk}=\delta^{i}_{~k}$~if~$h_{ij}(0),~h^{ij}(0)$~are
chosen as matrix inverses.\\
By (144), (146) one has
\begin{equation}\partial_{z}K_{[ij]}=-(K_{m}^{~m}+4Z^{-1})K_{[ij]}\end{equation}
so that by Theorem 1 of (Rendall and Schmidt 1991)~$K_{[ij]}\equiv 0$ and now (144) implies
that ~$K_{ij}=\frac{1}{2}\partial_{z}h_{ij}$. \\
Finally, by (147)-(148) there follows
\begin{equation}\partial_{z}(\bar{\zeta}-\zeta)=-(Z^{-1}+K)(\bar{\zeta}-\zeta)\end{equation}
so that~$\zeta=\bar{\zeta}$.
\\We conclude that for each Bianchi type 3-metric~$h_{ij}$~on~$G$~, there exists a unique
Bianchi type~$\gamma=1$~cosmology having an isotropic singularity,
with~$h_{ij}$~as initial metric in the chosen gauge. In particular, the Vanishing
Weyl Tensor conjecture goes through as
for~$1<\gamma\leq 2$.
\section{Concluding remarks}
The idea of constructing cosmological singularities by means of an infinite conformal transformation provides a mathematical framework for studying the Weyl Tensor Hypothesis of Penrose. In particular one may seek to generate cosmologies with isotropic singularities via a Cauchy problem with data given at the singularity surface. By allowing the conformal factor to be non-smooth at the singularity we have been able to solve this problem for perfect fluid models with equations of state other than that for 
radiation. For these other equations of state there are as many
cosmologies as for \linebreak[4] $\gamma=4/3$~and they are determined
by the intrinsic geometry of the singularity surface, with no free
data for the matter.

If one wishes to impose the condition~$C^a_{~bcd}=0$~initially then the uniqueness results obtained show that the spacetime geometry must be exactly FRW in a neighbourhood of the big-bang singularity.

One may now ask whether the picture painted above remains the same if
one considers other matter models. What, for example, happens to the
conjecture regarding vanishing of the Weyl tensor if we are able to
give some matter data at the singularity? Indeed, are there matter
models for which it is possible to give data at the singularity? This 
question will be addressed in a second paper in which we shall consider
the Cauchy problem for the conformal Einstein-Vlasov equations.

\appendix
\section{Explicit matrix representations for the $1<\gamma<2$ field
equations}
\subsection{Newman's formalism}
Here we present the formalism of Newman (1993b) which allows tensorial
evolution systems to be written in matrix form.
\\\\
Let~$V$~be an n-dimensional vector space with
dual~$V^{*}$. Let~$V$~be equipped with a basis,
and~$V^{*}$~with a dual
basis. Let~$h$~and~$h^{*}$~be covariant and
contravariant tensors over~$V$~of rank 2. For any
tensor~$T$~of total rank r over~$V$, with
components~$T^{i_{1}\ldots}_{~~~~~\ldots i_{r}}$~define
the~$n^{r}$~component row and column vectors
\begin{displaymath}(\mathbf{T})_{.}=(T^{1\ldots}_{~~~~\ldots
11},T^{1\ldots}_{~~~~\ldots 12},\ldots ,T^{1\ldots}_{~~~~\ldots
1n},T^{1\ldots}_{~~~~\ldots 21},\ldots,T^{n\ldots}_{~~~~\ldots
nn})\end{displaymath}
\begin{displaymath}(T)^{.}=\left(\begin{array}{ccccccc}T^{1\ldots}_{~~~~\ldots
11}\\
T^{1\ldots}_{~~~~\ldots 12}\\
\vdots\\
T^{1\ldots}_{~~~~\ldots 1n}\\
T^{1\ldots}_{~~~~\ldots 21}\\
\vdots\\
T^{n\ldots}_{~~~~\ldots nn}\end{array}\right)\end{displaymath}
A general matrix will be represented in the form~$M^{\cdot}_{~\cdot}$,
with~$M^{i}_{~\cdot}$~the ~$i^{\textrm{th}}$~row and~$M^{\cdot}_{~j}$~the
~$j^{\textrm{th}}$~column.. For any~$n^{k}\times n^{k}$~square
matrix~$M^{\cdot}_{~\cdot}$, and each~$i=1,2,\ldots
,n$~let~${}^{(i)}M^{\cdot}_{~\cdot}$~denote the matrices formed by
the~$1^{\textrm{st}},2^{\textrm{nd}},\ldots ,~n^{\textrm{th}}$~groups
of~$n^{k-1}$~rows of~$M^{\cdot}_{~\cdot}$~respectively, and
let~${}_{(i)}M^{~\cdot}_{~\cdot}$~be the matrices formed from the ~$1^{\textrm{st}},
2^{\textrm{nd}},\ldots , n^{\textrm{th}}$~groups of~$n^{k-1}$~columns
of~$M^{\cdot}_{~\cdot}$. Recall that the tensor product of a~$p\times
q$~matrix~$M^{\cdot}_{~\cdot}$~and an~$r\times s$~matrix~$N^{\cdot}_{~\cdot}$~may be
represented as a ~$pr\times qs$~matrix:
\begin{displaymath}M^{\cdot}_{~\cdot}\otimes
N^{\cdot}_{~\cdot}=\left(\begin{array}{cccc}M^{1}_{~1}N^{\cdot}_{~\cdot}&
M^{1}_{~2}N^{\cdot}_{~\cdot}&\cdots
& M^{1}_{~q}N^{\cdot}_{~\cdot}\\
M^{2}_{~1}N^{\cdot}_{~\cdot}& M^{2}_{~2}N^{\cdot}_{~\cdot}&\cdots &
M^{2}_{~q}N^{\cdot}_{~\cdot}\\
\vdots &\vdots &\ddots &\vdots\\
M^{p}_{~1}N^{\cdot}_{~\cdot}& M^{p}_{~2}N^{\cdot}_{~\cdot}&\cdots &
M^{p}_{~q}N^{\cdot}_{~\cdot}\end{array}\right)\end{displaymath}
Clearly this representation respects the associativity of tensor
products. For each~$k=1,2,3,\ldots$~one may therefore define the
~$n^{k}\times n^{k}$~matrix
\begin{displaymath}(\stackrel{k}{h})^{\cdot}_{~\cdot}=\underbrace{(h)^{\cdot}_{~\cdot}\otimes\ldots\otimes
(h)^{\cdot}_{~\cdot}}_{k~\textrm{factors}}
\end{displaymath}
where
\begin{displaymath}(h)^{\cdot}_{~\cdot}=\left(\begin{array}{cccc}h_{11}&
h_{12}& \cdots & h_{1n}\\
h_{21}& h_{22}&\cdots & h_{2n}\\
\vdots &\vdots &\ddots &\vdots\\
h_{n1}& h_{n2}& \cdots & h_{nn}\end{array}\right)\end{displaymath}
If~$h$~is a symmetric tensor
then~$(\stackrel{k}{h})^{\cdot}_{~\cdot}$~is a symmetric
matrix. Inductively one can show that if~$h$~and~$h^{*}$~are tensor
inverses in the sense of 
\begin{displaymath}h^{*ij}h_{jk}=\delta^{i}_{~k}\end{displaymath}
then~$(\stackrel{k}{h})^{\cdot}_{~\cdot}$~and~$(\stackrel{k}{h^{*}})^{\cdot}_{~\cdot}$~are
matrix inverses:
\begin{displaymath}\sum_{\sigma=1}^{n^{k}}(\stackrel{k}{h^{*}})^{\cdot}_{~\sigma}(\stackrel{k}{h})^{\sigma}_{~\cdot}=I^{n^{k}}\end{displaymath}
$(\stackrel{k}{h^{*}})^{\cdot}_{~\cdot}$~is defined in terms
of~$h^{*}$~by analogy with the preceding definition
of~$(\stackrel{k}{h})^{\cdot}_{~\cdot}$~in terms of ~$h$. Moreover,
letting~$h_{s}$~and~$h^{*}_{s}$~denote the symmetric parts
of~$h$~and~$h^{*}$~one may, for
each~$k=1,2,3,\ldots$~define~$(\stackrel{k}{h_{s}})^{\cdot}_{~\cdot}$~and~$(\stackrel{k}{h^{*}_{s}})^{\cdot}_{~\cdot}$.\\
Suppose~$h$~and~$h^{*}$~are tensor inverses and let~$T$~be any
contravariant tensor of rank~$r$~over~$V$. Then, defining a covariant
tensor~$T^{\flat}$~of rank~$r$~by
\begin{displaymath}T^{\flat}_{i_{1}\ldots i_{r}}=h_{i_{1}j_{1}}\ldots
h_{i_{r}j_{r}}T^{j_{1}\ldots j_{r}}\end{displaymath}
one has a matrix representation of the raising and lowering
operations:
\begin{displaymath}(T^{\flat})^{\cdot}=\sum_{\sigma=1}^{n^{r}}(\stackrel{r}{h})^{\cdot}_{~\sigma}(T)^{\sigma},~~~~(T)^{\cdot}=\sum_{\sigma=1}^{n^{r}}(\stackrel{r}{h^{*}})^{\cdot}_{~\sigma}(T^{\flat})^{\sigma}\end{displaymath}
There exists a unique symmetric~$n^{2}\times
n^{2}$~matrix~$J^{\cdot}_{~\cdot}$~such that, for every second rank
covariant tensor~$T_{ab}$~, the tensor~$\bar{T}_{ab}\equiv
T_{ba}$~satisfies
\begin{displaymath}(\bar{T})^{\cdot}=J^{\cdot}_{~\sigma}(T)^{\sigma},~~~~(\bar{T})_{~\cdot}=(T)_{\sigma}J^{\sigma}_{~\cdot}\end{displaymath}
In particular, for~$n=3$~one has
\begin{displaymath}J^{\cdot}_{~\cdot}=\left(\begin{array}{ccccccccc}1&0&0&0&0&0&0&0&0\\
0&0&0&1&0&0&0&0&0\\
0&0&0&0&0&0&1&0&0\\
0&1&0&0&0&0&0&0&0\\
0&0&0&0&1&0&0&0&0\\
0&0&0&0&0&0&0&1&0\\
0&0&1&0&0&0&0&0&0\\
0&0&0&0&0&1&0&0&0\\
0&0&0&0&0&0&0&0&1\end{array}\right)\end{displaymath}
$J^{\cdot}_{~\cdot}$~also plays an analogous role for second rank
contravariant tensors.\\
For each~$r=1,2,\ldots$~there exists a unique~$n^{r}\times
n^{r}$~matrix~$\stackrel{r}{E^{\cdot}_{~\cdot}}$~such that, for any
contravariant or covariant tensor field~$T$~of rank~$r$, one has
\begin{displaymath}(T_{s})^{\cdot}=\sum_{\sigma=1}^{n^{r}}\stackrel{r}{E^{\cdot}_{~\sigma}}(T)^{\sigma}\end{displaymath}
where~$T_{s}$~is the symmetric part of~$T$. In particular,
for~$n=3$~one has~$\stackrel{1}{E^{\cdot}_{~\cdot}}=I^{3}$~and
\begin{displaymath}\stackrel{2}{E^{\cdot}_{~\cdot}}=\left(\begin{array}{ccccccccc}1&0&0&0&0&0&0&0&0\\
0&\frac{1}{2}&0&\frac{1}{2}&0&0&0&0&0\\
0&0&\frac{1}{2}&0&0&0&\frac{1}{2}&0&0\\
0&\frac{1}{2}&0&\frac{1}{2}&0&0&0&0&0\\
0&0&0&0&1&0&0&0&0\\
0&0&0&0&0&\frac{1}{2}&0&\frac{1}{2}&0\\
0&0&\frac{1}{2}&0&0&0&\frac{1}{2}&0&0\\
0&0&0&0&0&\frac{1}{2}&0&\frac{1}{2}&0\\
0&0&0&0&0&0&0&0&1\end{array}\right)\end{displaymath}
\subsection{The reduced field equations for~$1<\gamma<2$}
If we define the 63-component column vector
\begin{displaymath}u=\left(\begin{array}{c}(h^{*})^{\cdot}\\
(h)^{\cdot}\\
V\\
\zeta\\
(\chi)^{\cdot}\\
(k)^{\cdot}\\
(\gamma)^{\cdot}\\
(\xi)^{\cdot}\\
\nu\end{array}\right)\end{displaymath}
then the ~$63\times 63$~matrices~$A^{\alpha}(u),~C(u)$~appearing in the reduced
equations (70) can be written\\
~~$A^{0}(u)=$
\begin{displaymath}\left(\begin{array}{ccccccccc}I^{9}&&&&&&&&\\
&I^{9}&&&&&&&\\
&& I^{1}&&&&& O        &\\
&&& I^{1}&&&&&\\
&&&& -(h^{*}_{s})^{\cdot}_{~\cdot}&&&&\\
&&&&& V^{2}(\stackrel{2}{h_{s}})^{\cdot}_{~\cdot}&&&\\
&& O &&&& -(\stackrel{3}{h^{*}_{s}})^{\cdot}_{~\cdot}&&\\
&&&&&&&\left(\frac{2-3\gamma}{2\gamma}\right)(h^{*}_{s})^{\cdot}_{~\cdot}&\\
&&&&&&&&\frac{V^{2}}{(2-\gamma)}I^{1}
\end{array}\right)\end{displaymath}
~~$A^{i}(u)=$
\begin{displaymath}\left(\begin{array}{ccccccccc}0&&&&&&&&\\
&0&&&\ldots &&&&\\
&&0&&&&&&\\
&&&0&&&&&\\
&\vdots&&&0&0&0&0&-\frac{F}{E}(h^{*}_{s})^{\cdot}_{~i}\\
&&&&0&0&
(h^{*}_{s})^{i}_{~\cdot}\otimes\stackrel{2}{E^{\cdot}_{~\cdot}}&
-\frac{2}{G}\stackrel{2}{{}_{(i)}E^{\cdot}_{~\cdot}}&0\\
&&&&0&
(h^{*}_{s})^{\cdot}_{~i}\otimes\stackrel{2}{E^{\cdot}_{~\cdot}}&0&0&0\\
&&&&0&
-\frac{2}{G}\stackrel{2}{{}^{(i)}E^{\cdot}_{~\cdot}}&0&0&\frac{3G}{2F}(h^{*}_{s})^{\cdot}_{~i}\\
&&&&
-\frac{F}{E}(h^{*}_{s})^{i}_{~\cdot}&0&0&\frac{3G}{2F}(h^{*}_{s})^{i}_{~\cdot}&0\end{array}\right)\end{displaymath}
~~$C(u)=$
\begin{displaymath}\left(\begin{array}{ccccccccc}0&&&&&&&&\\
&0&&\ldots &&&&&\\
&&0&&&&&&\\
&\vdots &&
-I^{1}&0&D_{1}(h_{s})_{\cdot}&0&0&D_{2}I^{1}\\
&&&0&D_{3}(h^{*}_{s})^{\cdot}_{~\cdot}&0&0&D_{4}(h^{*}_{s})^{\cdot}_{~\cdot}&0\\
&&&-2V^{2}(h_{s})^{\cdot}&0&D_{5}(-2(\stackrel{2}{h_{s}})^{\cdot}_{~\cdot}+(h_{s})^{\cdot}\otimes
(h_{s})_{\cdot})&0&0&D_{6}(h_{s})^{\cdot}\\
&&&0&0&0&0&0&0\\
&&&0&D_{7}(h^{*}_{s})^{\cdot}_{~\cdot}&0&0&D_{8}(h^{*}_{s})^{\cdot}_{~\cdot}&0\\
&&&D_{9}I^{1}&0&D_{10}(h_{s})_{\cdot}&0&0&D_{11}I^{1}\end{array}\right)\end{displaymath}
where
\begin{displaymath}D_{1}=\frac{3(2-\gamma)^{2}}{(3\gamma-2)^{2}(1-5\gamma)}~~~~D_{2}=\frac{30\gamma(\gamma-2)F^{-1}}{(3\gamma-2)^{2}(1-5\gamma)}\end{displaymath}
\begin{displaymath}D_{3}=\frac{-4P^{-1}}{(3\gamma-2)}~~~~D_{4}=\frac{2EG^{-1}}{P^{2}(3\gamma-2)}\end{displaymath}
\begin{displaymath}D_{5}=\frac{2V^{2}}{(3\gamma-2)}~~~~D_{6}=\frac{-4V^{2}(\gamma-3)F^{-1}}{(3\gamma-2)(\gamma-2)}\end{displaymath}
\begin{displaymath}D_{7}=\frac{8GE^{-1}}{(2-3\gamma)}~~~~D_{8}=\frac{4P^{-1}}{(3\gamma-2)}\end{displaymath}
\begin{displaymath}D_{9}=\frac{V^{2}F(3\gamma-2)}{(2-\gamma)}~~~D_{10}=\frac{V^{2}F}{(3\gamma-2)}~~~D_{11}=\frac{2V^{2}(3\gamma-5)}{(2-\gamma)(3\gamma-2)}\end{displaymath}
\subsection{The recovery equations for~$1<\gamma<2$}
Define the 24-component column vector
\begin{displaymath}v=\left(\begin{array}{cccccccc}(\pi)^{\cdot}\\
\tau\\(\alpha)^{\cdot}\\(\beta)^{\cdot}\\(\omega)^{\cdot}\\\delta\\\eta\\\kappa\end{array}\right)\end{displaymath}
then the matrices ~$a^{\alpha}(v),~c(v)$~appearing in the recovery
equations (105) are as follows\\
~$a^{0}(v)=$
\begin{displaymath}\left(\begin{array}{cccccccc}I^{3}&&&&&&&\\
&I^{1}&&&&&&\\&&(\stackrel{2}{h^{-1}})^{\cdot}_{~\sigma}\left(\frac{1}{2}I^{9}+C_{3}J\right)^{\sigma}_{~\cdot}&&&&O&\\
&&&C_{4}(h^{-1})^{\cdot}_{~\cdot}&&&&\\
&&&&-V^{2}(h^{-1})^{\cdot}_{~\cdot}&&&\\
&&&&&C_{5}I^{1}&&\\
&O&&&&&-V^{2}(h^{-1})^{\cdot}_{~\cdot}&\\
&&&&&&&V^{4}I^{1}\end{array}\right)\end{displaymath}
where
\begin{displaymath}C_{3}=\frac{\gamma-1}{8\gamma}~~~~C_{4}=\frac{V^{2}(2-3\gamma)}{4\gamma}\end{displaymath}
\begin{displaymath}C_{5}=\frac{V^{2}}{4}\left(\frac{5\gamma-1}{3\gamma-1}\right)\end{displaymath}
while~$a^{i}(v)=$
\begin{displaymath}\left(\begin{array}{cccccccc}0&\ldots&&&&&&\\
\vdots&0&\ldots&&&&&\\
&\vdots&0&Q^{\cdot}_{~\sigma}({}_{(i)}\stackrel{2}{h^{-1}})^{\sigma}_{~\cdot}&0&0&&\\
&&({}^{(i)}\stackrel{2}{h^{-1}})^{\cdot}_{~\sigma}Q^{\sigma}_{~\cdot}&0&0&C_{6}(h^{-1})^{\cdot}_{~i}&&\\
&&0&0&0&C_{7}(h^{-1})^{\cdot}_{~i}&&\\
&&0&C_{6}(h^{-1})^{i}_{~\cdot}&C_{7}(h^{-1})^{i}_{~\cdot}&0&\vdots&\\
&&&&&\ldots&0&\vdots\\
&&&&&&\ldots&0\end{array}\right)\end{displaymath}
where
\begin{displaymath}Q=\frac{1}{2}I^{9}+C_{3}J\end{displaymath}
\begin{displaymath}C_{6}=\frac{(3-\gamma)V^{2}}{4(3\gamma-1)C_{2}}~~~~C_{7}=\frac{V^{2}}{2}\left(\frac{2\gamma+1}{1-3\gamma}\right)\frac{C_{2}}{C_{1}}\end{displaymath}
and~$c(v)=$
\begin{displaymath}\left(\begin{array}{cccccccc}0&\ldots&&&&&&\\
\vdots&0&\ldots&&&&&\\
&&0&\ldots&&&&\\
&&\vdots&C_{8}(h^{-1})^{\cdot}_{~\cdot}&C_{9}(h^{-1})^{\cdot}_{~\cdot}&0&C_{10}(h^{-1})^{\cdot}_{~\cdot}&0\\
&&&0&0&0&0&0\\
&&&0&0&0&0&C_{11}I^{1}\\
&&&-V^{2}(h^{-1})^{\cdot}_{~\cdot}&0&0&V^{2}(h^{-1})^{\cdot}_{~\cdot}&0\\
&&&0&0&\frac{V^{4}}{C_{2}}I^{1}&0&-V^{4}I^{1}\end{array}\right)\end{displaymath}
where
\begin{displaymath}C_{8}=\frac{V^{2}(4\gamma-3)}{(3\gamma-2)\gamma}~~~~C_{9}=\frac{2V^{2}(4\gamma-3)}{C_{1}(3\gamma-2)(1-3\gamma)}\end{displaymath}
\begin{displaymath}
 C_{10}=\frac{3V^{2}(2-\gamma)(\gamma-1)}{(3\gamma-2)^{2}}~~~C_{11}=V^{4}\left\{\frac{2(13\gamma-8)+6(1-V^{P})(5\gamma
^
{2}-4\gamma+3)}{(1-3\gamma)(3\gamma-2)^{2}}\right\}\end{displaymath}
and~$C_{1},~C_{2}$~were defined in section 4.4.

\section*{References}
\begin{description}
\item R. A. Adams 1975, \emph{Sobolev spaces} (New York:
Academic Press)
\item C. M. Claudel and K.P. Newman 1998, \emph{Proc. R. Soc. Lond.} 454, 1073-1107
\item S. W. Goode 1987, \emph{Gen. Rel. Grav.} 19, 1075-1082
\item S. W. Goode and J. Wainwright 1985,
\emph{Class. Quant. Grav.} 2, 99-115
\item T. Kato 1975, \emph{Arch. Ration. Mech. Anal.} 58, 181-205
\item R. P. A. C. Newman 1993a, \emph{Proc. R. Soc. Lond.} 443, 473-492
\item R. P. A. C. Newman 1993b, \emph{Proc. R. Soc. Lond.} 443, 493-515
\item R. Penrose 1979, in \emph{General Relativity: an Einstein
centennial volume}~ed. W. Israel and S. W. Hawking (Cambridge: CUP)
\item R. Penrose 1990, \emph{The Emperor's New Mind} (Oxford:
OUP)
\item R.Racke 1992, \emph{Lectures on nonlinear evolution equations, Aspects of Mathematics} vol.E19 (Vieweg)
\item A.D.Rendall and B.G.Schmidt 1991, \emph{Class. Quant. Grav.} 8, 985-1000
\item M. E. Taylor 1991, \emph{Pseudodifferential operators
and nonlinear PDE,}\linebreak[4]\emph{Progress in Mathematics} vol. 100 (Birkh\"auser)
\item K. P. Tod 1987, \emph{Class. Quant. Grav.} 4, 1457-146
\item K. P. Tod 1990, \emph{Class. Quant. Grav.} 7, L13-L16
\item K. P. Tod 1991, \emph{Class. Quant. Grav.} 8, L77-L82
\item K. P. Tod 1992, \emph{Rend. Sem. Mat. Univ. Pol. Torino}
vol. 50, 1, 69-92
\item R. M. Wald 1984, \emph{General Relativity}
(University of Chicago Press)
\end{description}
\end{document}